\documentclass[aps, prl, reprint,longbibliography]{revtex4-1}

\usepackage{graphicx}
\usepackage{hyperref}
\hypersetup{
    bookmarks=false,         
    unicode=false,          
    pdftoolbar=false,        
    pdfmenubar=true,        
    pdffitwindow=false,     
    pdfstartview={FitH},    
    pdftitle={},    
    pdfauthor={},     
    pdfsubject={},   
    pdfcreator={},   
    pdfproducer={}, 
    pdfkeywords={quantum scars} {non-ergodic dynamics}, 
    pdfnewwindow=true,      
    colorlinks=true,       
    linkcolor=black,          
    citecolor=magenta,        
    filecolor=magenta,      
    urlcolor=magenta
}

\usepackage[normalem]{ulem}
\PassOptionsToPackage{usenames,dvipsnames}{xcolor}
\usepackage{color}
\usepackage{bm}
\usepackage{amsmath, amsthm, amssymb}
\usepackage{mathtools, xfrac}

\renewcommand{\and}{\quad \text{and}\quad}
\newcommand{\ie}{i.\,e. }

\newcommand{\eg}{e.\,g. }

\renewcommand{\vec}{\mathbf}
\renewcommand{\d}{\mathrm{d}}

\newcommand{\im}{\mathrm{i}\mkern1mu}

\begin{document}
\title{{Topological charges of} periodically kicked molecules}

\author{Volker Karle}
\email{volker.karle@ist.ac.at}

\author{Areg Ghazaryan}
\email{areg.ghazaryan@ist.ac.at}

\author{Mikhail Lemeshko}
\email{mikhail.lemeshko@ist.ac.at}
\affiliation{Institute of Science and Technology Austria, Am Campus 1, 3400 Klosterneuburg, Austria}
\date{\today}
\begin{abstract}
We show that the simplest of existing molecules -- closed-shell diatomics not interacting with one another -- host {topological charges} when driven by periodic far-off-resonant laser pulses. A periodically kicked molecular rotor can be mapped onto a ``crystalline'' lattice in angular momentum space. This allows to define quasimomenta and the band structure in the Floquet representation, by analogy with the Bloch waves of solid-state physics. Applying laser pulses spaced by $1/3$ of
the molecular rotational period creates a lattice with three atoms per unit cell with staggered hopping. {Within the synthetic dimension of the laser strength, we discover Dirac cones with topological charges}. These Dirac cones, topologically protected by reflection and time-reversal symmetry, are reminiscent  of (although not equivalent to) that seen in graphene. They -- and the corresponding edge states -- are broadly tunable by adjusting the laser {strength} and can be  observed in present-day experiments by measuring molecular alignment and populations of rotational levels. This paves the way to study controllable topological physics in  gas-phase experiments with small molecules as well as to classify dynamical molecular states by their topological invariants.
\end{abstract}

\maketitle
The quantum nature of electrons in solids gives rise to a number of fascinating phenomena, such as the quantum Hall effect, that are ultimately related to the geometric and topological properties of the Brillouin zone~\cite{girvin_modern_2019}. {These phenomena are characterized by \textit{topological charges} -- a type of a quantum number that describes the topology of the system. These topological phases} show unique properties, such as  quantized transport and the bulk-boundary correspondence~\cite{bernevig_topological_2013}. The field of topological phases has greatly expanded since the works of Thouless~\cite{thouless_quantized_1982} and Haldane~\cite{haldane_nonlinear_1983,haldane_model_1988}, leading to discoveries of several new phases, such as topological insulators, Weyl semimetals, and  topological superconductors~\cite{qi_topological_2011,armitage_weyl_2018,sato_topological_2017}. 

Unlike the translational motion of an electron in a lattice, rotations of a molecule correspond to the non-Abelian group $\mathrm{SO(3)}$. While free rotations basically correspond to trivial paths in that manifold \cite{balakrishnan2020mathematical,khatua2022berry}, in this Letter we show that laser pulses can guide the molecule along topologically nontrivial paths, allowing for nonzero Berry phases and alike. In particular, we explore the similarities between the Bloch theorem (for a system periodic in space) and the Floquet theorem (for a system periodic in time) to show that a molecule driven by periodic laser pulses  can be mapped onto a  translationally invariant hopping model hosting nontrivial topology.
Although the ideas of topology in molecules have been extensively exploited in the context of conical intersections of potential energy surfaces in real space~\cite{ConicalBook, RyabinkinACR17} (including light-induced conical intersections~\cite{Moiseyev_2008, Hal_sz_2011}), the results presented here use {a novel approach} and allow to directly bridge the ideas of {symmetry}-protected phases in condensed-matter physics with the realm of molecules.

At energies well below electronic and vibrational excitations diatomic molecules essentially behave as rigid linear rotors~\cite{LevebvreBrionField2}. A two-dimensional kicked rotor is a paradigmatic model used to study nonlinear dynamics, dynamical localization and quantum chaos~\cite{Santhanam_2022, casati_quantum_2006}. Since the original works of Casati and Chirikov~\cite{casati_stochastic_1979, shepelyanskii1981dynamical,fishman1982chaos, casati_anderson_1989}, the predictions of the theory have been verified in several experiments, e.g., with atoms in microwave fields~\cite{graham_dynamical_1992, moore_observation_1994}, Rydberg atoms~\cite{blumel_dynamical_1991}, atomic matter waves~\cite{chabe2008experimental}, and Bose-Einstein condensates~\cite{WeldRotorBEC21}. Topological aspects of double kicked two-dimensional rotors have also been extensively explored in connection to spectra resembling the Hofstadter butterfly and the corresponding Chern numbers~\cite{wang2008proposal,zhou2018recipe,ho2012quantized}. Periodically driven three-dimensional molecular rotors have been studied theoretically with a particular focus on resonances and Anderson localization~\cite{BluemelJCP86, FlossPRA12, flos_anderson_2013} and edge states~\cite{FlossPRE15, BakmanEPJB19}. Several phenomena, such as quantum resonances~\cite{ZhdanovichPRL12}, Bloch oscillations~\cite{FlossPRL15, flos_anderson_2014} and dynamical localization~\cite{bitter_experimental_2016, bitter_experimental_2017, bitter_control_2017} have already been observed in experiments with molecules. Moreover, recent advances in imaging of molecular rotational dynamics~\cite{bert_optical_2020,lin_visualizing_2015, karamatskos_molecular_2019} and control of their angular degrees of freedom~\cite{koch_quantum_2019, MitraPRA22} open new possibilities to probe kicked rotor physics.
While these advances are particularly important for the understanding of reactions and
other fundamental processes in physical chemistry~\cite{zare_laser_1998, xie_selective_2014}, they can -- as we show below -- find applications in other seemingly unrelated areas of physics, such as the study of topological phases.

In this Letter we demonstrate that apart from the rich physics related to transport and localization, driving even the simplest molecules by specifically designed periodic laser pulses allows to probe the nontrivial topology of their rotational states. {This allows to apply the vocabulary and theorems developed for topological materials in the realm of chemical physics, and to potentially engineer topologically protected molecular states with new chemical applications.} Here we engineer an effective topological semimetal with linear dispersing topological edge states, however, more involved models characterized by other topological invariants can potentially be realized as well. In addition to a higher degree of control achievable in experiment, periodically kicked molecules are able to form multiband topological systems. This paves the way to realize non-Abelian topological phases, whose study in solid-state settings has been quite limited so far.

In what follows, we consider the simplest case of a linear closed-shell molecule which is periodically kicked  by a far-off-resonant, linearly polarized laser. The general idea, however, is straightforward to extend to more complex molecules (\eg symmetric and asymmetric tops) and to other kinds of fields, which might further expand the range of realizable Hamiltonians. When the laser pulse duration is significantly shorter than the rotational period of the molecule~\footnote{We provide a justification of the sudden approximation in the Supplementary Material~\cite{sup}, which includes Ref.~\cite{dion2001orienting,dion_laser-induced_1999,seideman1999revival,dion2002optimal, mirahmadi2018dynamics,rauch2006time,arkhipov2022half,barth2008nonadiabatic,ChatterleyPRL20}.}, we can write the
Hamiltonian {(in units of $\hbar \equiv 1$)} as follows~\cite{averbukh2001angular,bitter_control_2017}:
\begin{equation}
    \label{eq:H_full}
    \begin{aligned}
        \hat{H}_\text{mol}(t) &= B \hat{\mathbf{L}}^2 - \underbrace{ \left [ P_1\cos^2(\hat{\theta}) + P_2 \cos(\hat{\theta}) \right]}_{\equiv \hat{V}(P_1,P_2)} \sum_{q=0}^{\infty} \delta(t - q T)
    \end{aligned}
\end{equation}
Here {$B = \pi /\tau_B$} is the molecular rotational constant with $\tau_B$ the rotational period, {$\hat{\mathbf{L}}$ is the angular momentum operator, with centrifugal distortion neglected~\footnote{For some molecules, the centrifugal distortion constant $D$ is not negligible and the $D\,l^2(l+1)^2$ term has to be added. This term becomes important for larger values of $l$ and can be neglected in our discussion for now. Practically, it will introduce a second boundary with some $l_\mathrm{max}$ (the so-called ``Anderson wall''), as was shown in~\cite{flos_anderson_2014}\label{note1}.}}. The laser pulses are defined by their strengths, $P_1,P_2$, and the time period between them, $T$. {$\hat{\theta}$ gives the angle between the linear polarization of both lasers and the molecular axis.} Unlike most other models taking into account either the $\cos^2(\hat \theta)$ term (``regular'' multi-cycle infrared pulses) or the  $\cos(\hat \theta)$ term (terahertz half- or few-cycle pulses~\cite{you1993generation,barth2008nonadiabatic,dion2001orienting,averbukh2001angular,gershnabel_orientation_2006}), here we include both,
which allows to create tunable Dirac cones, not observable with either one of the terms~\cite{sup}. {For simplicity, we combined the two laser pulses in one potential, but the same behavior can be observed for two consecutive pulses. Both kind of pulses have been realized independently in experiments and used for orientation and alignment of molecules~\cite{karamatskos2019molecular,nevo2009laser,bert_optical_2020,lin_visualizing_2015, karamatskos_molecular_2019,cheng2019field}.}

Casati~\textit{et al.}~\cite{casati_stochastic_1979, casati_quantum_2006} have shown that a periodically driven pendulum, and hence also a molecule, displays two regimes: the \textit{dynamical localization regime}, \ie localization on a lattice in time (instead of a lattice in space), and the \textit{quantum resonance regime}, where the pendulum delocalizes. In this work, in order to achieve a banded system, we focus on the resonant regime~\cite{ZhdanovichPRL12,wimberger2014nonlinear}, \ie~$T =
\tau_B/N$. The time-translation operator of the Hamiltonian defined in Eq.~\eqref{eq:H_full} after one {driving period} takes the form
\begin{equation}
\label{eq:U}
    \hat{U} = \underbrace{e^{- \pi \im \hat{\mathbf{L}}^2/ N}}_{\text{Free rotation}}\underbrace{e^{\im \hat{V}(P_1,P_2)}}_{\text{Kick}}.
\end{equation}
The ideas described below are based on the analogy between ``real'' solid-state systems, which are periodic in space, and molecules periodically kicked in time. Periodically driven systems can be described in terms of the so-called Floquet states, $|\psi_n\rangle$, and the corresponding quasienergies, $\epsilon_n$~\cite{grifoni_driven_1998,breuer2002theory}.  These can be defined through the time-translation operator of Eq.~\eqref{eq:U} as $\hat{U}|\psi_n \rangle = e^{\im\epsilon_n}|\psi_n
\rangle$, or, equivalently, through the effective Hamiltonian $\hat{H} = \im \log \hat{U}$ with $\hat{H}|\psi_n \rangle = \epsilon_n|\psi_n \rangle$. At a quantum resonance, the quasienergies form $N$ bands {if $N$ is odd (for even $N$ the periodicity is $2N$ and there are $2N$ bands, see~\cite{sup})}, which is a result of the $N-$periodicity of the $e^{- \pi \im \hat{\mathbf{L}}^2/ N}$ operator. 

It is important to note that characterizing periodically kicked molecules in terms of their Floquet states and quasienergies has been previously done in a number of works, see e.g.~Refs.~\cite{flos_anderson_2014, FlossPRE15}. In what follows, we go one step further and demonstrate that at a quantum resonance one can introduce quasimomenta of periodically driven molecular states.
This makes it possible to work with ``molecular Bloch bands'' and to study their topology, which provides a direct bridge to the condensed matter systems.
\begin{figure}[t!]
    \centering
    \includegraphics[width=0.96\linewidth]{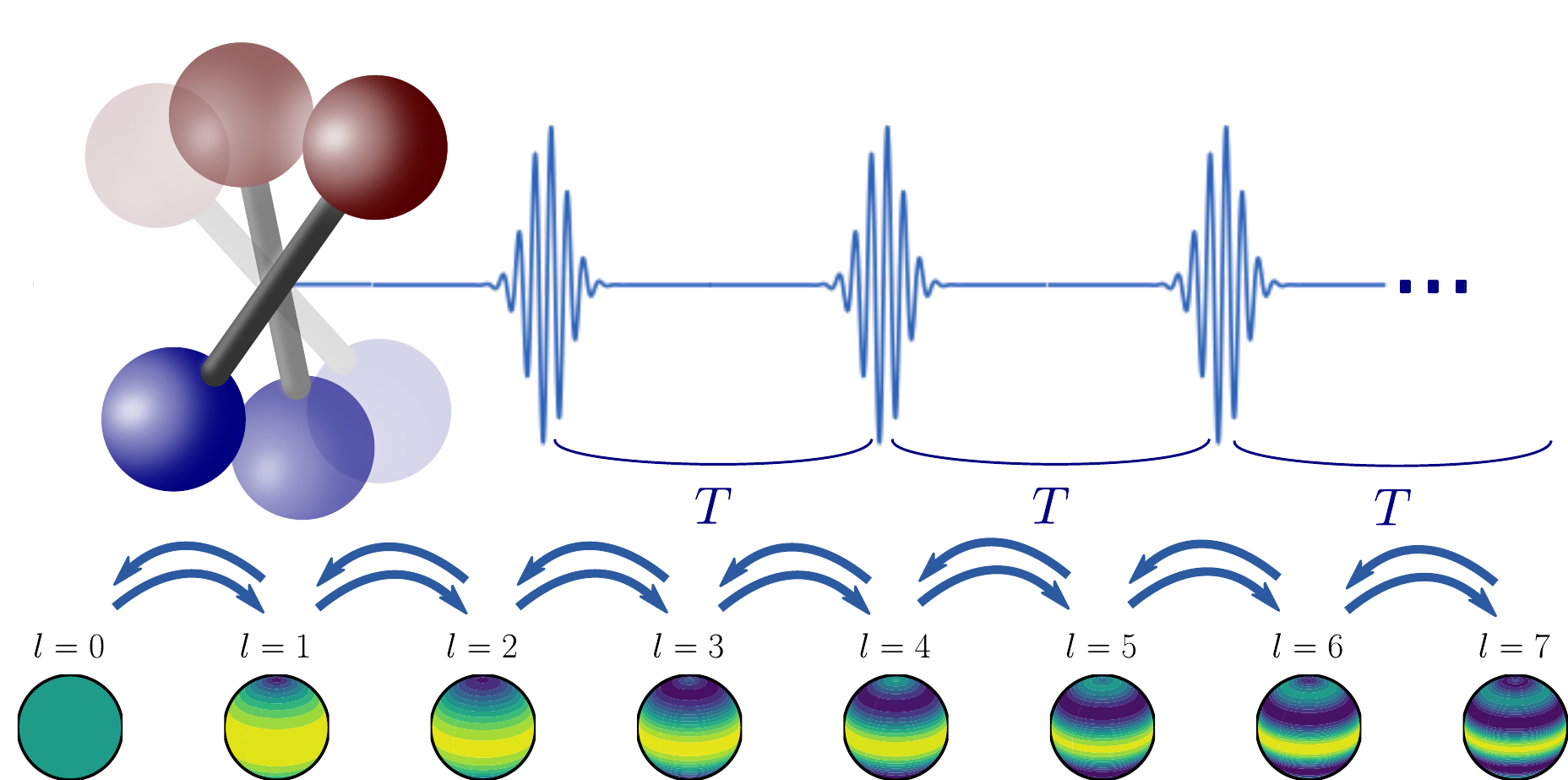}
    \caption{Illustration of the angular momentum lattice with the spherical harmonics of the molecule and the hopping between different lattice sites due to the periodic laser pulses, cf.\ Eq.~\eqref{eq:H_full}. For $l \gg 0$, the hopping terms converge to a constant and the lattice becomes translationally invariant. {The hopping terms are complex numbers given by the time-translation operator (see Fig.~S4 in the supplementary material~\cite{sup}).}}%
    \label{fig:fig1}
\end{figure}

\begin{figure*}[t]
	\centering
	\includegraphics[width=1.0\linewidth]{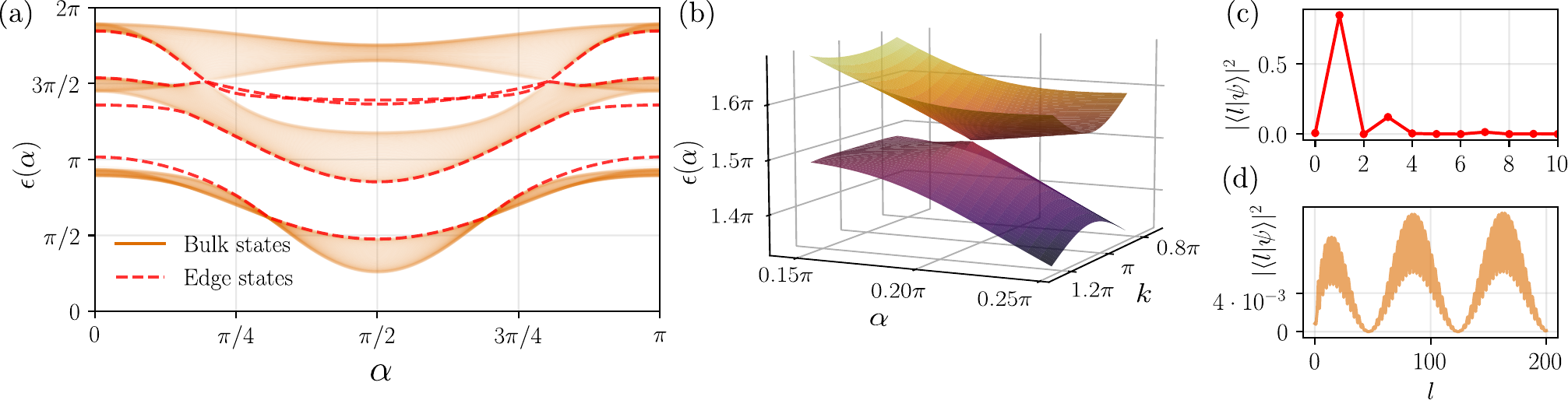}
	\caption{Results of exact diagonalization for $N=3$ and $P=2.5$, see Eq.~\eqref{eq:model1}.  (a) The full spectrum with the bulk states shown in orange and the edge states in red. Two linearly dispersing edge states (one for each edge) connect the Dirac cones. (b) The quasienergies $\epsilon_k$ near the Dirac cone, shown as a function of $k$ and $\alpha$. As a result of  time-reversal and reflection symmetries, the two bands touch at $k=\pi$. The Dirac cone is topologically protected and cannot be gapped out by any deformation which preserves the symmetry, {unless there is a merging transition where a positively charged cone annihilates a negatively charged one.} (c)  The absolute value of the wavefunction of the edge state. (d) A generic bulk state.}%
	\label{fig:Fig2}
\end{figure*}
In particular, we make use of the fact that for nonzero values of angular momentum, the ``hopping'' matrix elements between different angular momentum ``lattice sites'' converges to an approximately constant value, $H_{l,l'} \approx H_{l+N,l'+N}$ for $l,l' \gg 0$, with $H_{l,l'} = \langle l \vert \hat H \vert l' \rangle$, cf. Eq.~\eqref{eq:H_full}~\footnote{We demonstrate this behavior in the supplementary material~\cite{sup}, including~\cite{varshalovich1988quantum,flude_edmonds_1998}.}. This allows to create an effective translational invariant tight-binding model where the hopping is controlled by the periodic laser pulses, see Fig.~\ref{fig:fig1}. Note that while translational invariance is exact for a planar rotor, it is only approximately achieved for 3D molecular rotors in the limit of $l,l' \gg 0$, which, however, suffices for the purposes of our proposal.
{Let us assume that $l \in \mathbb{Z}$, then, the periodicity of the lattice allows us to define the Fourier transform with quasimomentum $k \in [0,2\pi)$ by
\begin{equation}
\label{eq:Heff}
    \underbrace{H_{ij}(k)}_{N\text{x}N \text{Matrix}} = \sum_{\Delta n} e^{-\im \Delta n \cdot k} H_{ij}(\Delta n)\\
\end{equation}
with $H_{ij}(\Delta n) = H_{i,N \cdot \Delta n + j}$ (in practice, we need to choose a cut-off for a finite system, see~\cite{sup}. Note that this finite-size Fourier transform is only approximate and does not capture the physics at the boundary. However, as it turns out, the spectrum obtained within this approximation agrees very well with the results of exact diagonalization and captures the behavior at a quantum resonance, \ie the indefinite growth of rotational energy).} Similarly, for any function in that space we define \begin{equation}
   f_i(k) = \sum_{\Delta n} e^{-\im \Delta n \cdot k} f(i + N\Delta n).
   \label{eq:fourier}
\end{equation}
Correspondingly, $H(k)\psi_n(k) = \epsilon_n(k)\psi_n(k)$ where $k$ plays the role of quasimomentum  of the angular momentum lattice and $\psi_n(k) \in \mathbb{C}^N$ (note that $H(k)$ and other operators are $N\times N$ matrices in $k$-space).

Since there are $N$ bands, the Hamiltonian can be written as $H(k) = a(k)\mathbb{I}_{N} +\sum_i d_i(k)\cdot\lambda_i$ in terms of the generalized Gell-Mann matrices $\lambda_i \in \mathrm{SU}(N)$, which form a linearly independent basis of traceless $N \times N$ matrices, and $a(k)\in \mathbb{R},\vec{d}(k) \in \mathbb{R}^{N^2-1}$, see Ref.~\cite{kauffman2007mathematics}. 
The mapping described above connects kicked molecules to single-particle models of condensed matter physics. Furthermore, the symmetries of the effective Hamiltonian are determined by the symmetries of the time-translation operator. Unlike a Hermitian system, however, this system exhibits periodic quasienergies; the free choice of the initial time results in a gauge freedom of the quasienergies. Since the first and the last band can touch over the periodic boundary, this opens up new possibilities such as anomalous topological Floquet insulators~\cite{lindner2011floquet,kitagawa2010topological}, which are out of the scope of this Letter.

Here we focus on the case of $N=3$ for intermediate coupling where no closing of all gaps occurs~\cite{FlossPRE15}, since this, to the best of our knowledge, is the simplest situation featuring nontrivial topological physics for a kicked molecule. We choose the following parametrization of the laser strength parameters:
\begin{equation}
P_1(\alpha)=P\cos^2(\alpha), \,P_2(\alpha)= P\sin^2(\alpha)
\label{eq:model1}
\end{equation}
in terms of the offset parameter $\alpha \in [0,2\pi)$, which serves as an effective second dimension in addition to quasimomentum and laser strength $P$. Since the linearly-polarized laser of Eq.~\eqref{eq:H_full} preserves the projection of angular momentum, $m$, we consider only the states with $m=0$.
Fig.~\ref{fig:Fig2} shows the results of exact diagonalization.  
There are three bands, shown in Fig.~\ref{fig:Fig2}(a), with the second and third touching at two points, forming Dirac cones, Fig.~\ref{fig:Fig2}(b). The Fourier transform shows that the bands touch exactly at $k=\pi$, indicating a symmetry in quasimomentum space. Indeed, the system has a reflection and a time-reversal symmetry that commute with each other~\cite{sup}, \ie
\begin{equation}
\mathcal{T}H_k = H^*_{-k}\mathcal{T} \quad\text{and}\quad\mathcal{R}H_k = H_{-k}\mathcal{R}.
\end{equation}
This implies that for the $k\rightarrow -k$ invariant points, i.e. $k=0$ or $k=\pi$, the Hamiltonian commutes with $\mathcal{T}$ and $\mathcal{R}$. From the 10-fold classification with crystal symmetries the class is $\mathrm{AI}_+\mathcal{R}_+$~\cite{chiu_classification_2016}. The topological analysis of the two Dirac cones proceeds as follows. 
We combine the two variables into a vector $\vec{k}=(k,\alpha),$ and compute the Berry connection, $\mathcal{A}_n(\vec{k}) = i\langle \psi_n(\vec{k})|\nabla_\vec{k}\psi_n(\vec{k})\rangle$, which is well-defined if we choose a gauge such that the eigenstates form a smooth manifold. Numerically evaluating the Berry phase~\cite{fukui2005chern} around one of these cones in the second or third band results in $\int_\gamma \mathcal{A}(\vec{k})_n \d\vec{k}=  \pm \pi$ thus proving the topological nature of these cones~\footnote{Another possibility is to  calculate the mirror invariants, which can be computed from the difference in the number of reflection-symmetric eigenstates at $k=0$ (or $k=\pi$) before and after the cone.}. As with spinless graphene, we observe two linear dispersing edge states, connecting the two Dirac cones. This is a result of the bulk-boundary correspondence for Dirac semimetals with crystal symmetries. For the bulk states we find plane-wave solutions as predicted by the Fourier-transformed Hamiltonian.  The edge states manifest themselves as states localized at the lower {(or upper~\footnote{From the rigid rotor approximation, the lattice in $l$ is assumed to be half-infinite. However, that is only an approximation, see footnote [50] and~\cite{flos_anderson_2014}. The centrifugal distortion leads to an effective upper bound $l_\mathrm{max}$, where another edge state appears.})} end of the lattice. Topologically non-protected edge states in kicked molecules have already been theoretically described~\cite{FlossPRE15, BakmanEPJB19}. The localized edge  states shown in Fig.~\ref{fig:Fig2}(c) are, on the other hand, topological, in the sense that they cannot be destroyed without merging the two Dirac cones.

The  topological characteristics of the band structure shown in Fig.~\ref{fig:Fig2} have observable experimental signatures. In order to demonstrate that, we consider the time evolution of a molecule adiabatically driven through the Dirac cones (for fast quenches the phenomenon persists, but is superimposed by the quench dynamics). We adiabatically change the parameter $\alpha$ of Eq.~\eqref{eq:model1} as 
\begin{equation}
   \alpha(t) = t \cdot (\alpha_c/t_c) 
   \label{eq:alpha}
\end{equation}
where $\alpha_c$ denotes the critical value of $\alpha$ for which the cones appear at $k_c=\pi$. One can choose $t_c$ as the time when one wishes the cones to emerge, here we set $t_c/T = 15$. Experimentally that protocol implies changing $\alpha$ after every kick. 
The Dirac cones form vortices in the $\mathbf{k}$-plane with singularities at $(k_c, \alpha_c)$. We demonstrate that the vortices lead to a flip of the molecular orientation during the time evolution even if we time-evolve a generic, experimentally realizable wave packet, populated in one of the bands involved in the crossing.

To this end, we consider a Gaussian wavepacket $\langle l|\phi(t=0)\rangle \propto e^{-(l-l_0)^2/2\Delta l^2}$ and project this state into the third band with a peak at $k=\pi$~\footnote{quasimomentum $k=\pi$ implies that the sign of the matrix elements changes every three angular momentum sites.}. This furnishes a state which is both localized in angular momentum $l$ and quasimomentum $k$. The occupation in the third band leads to high orientation and alignment signals. Henceforth we time-evolve this wavepacket according to the adiabatic protocol in real space and without approximation of Fourier space conversion. 

\begin{figure}[htpb] \centering
    \includegraphics[width=0.89\linewidth]{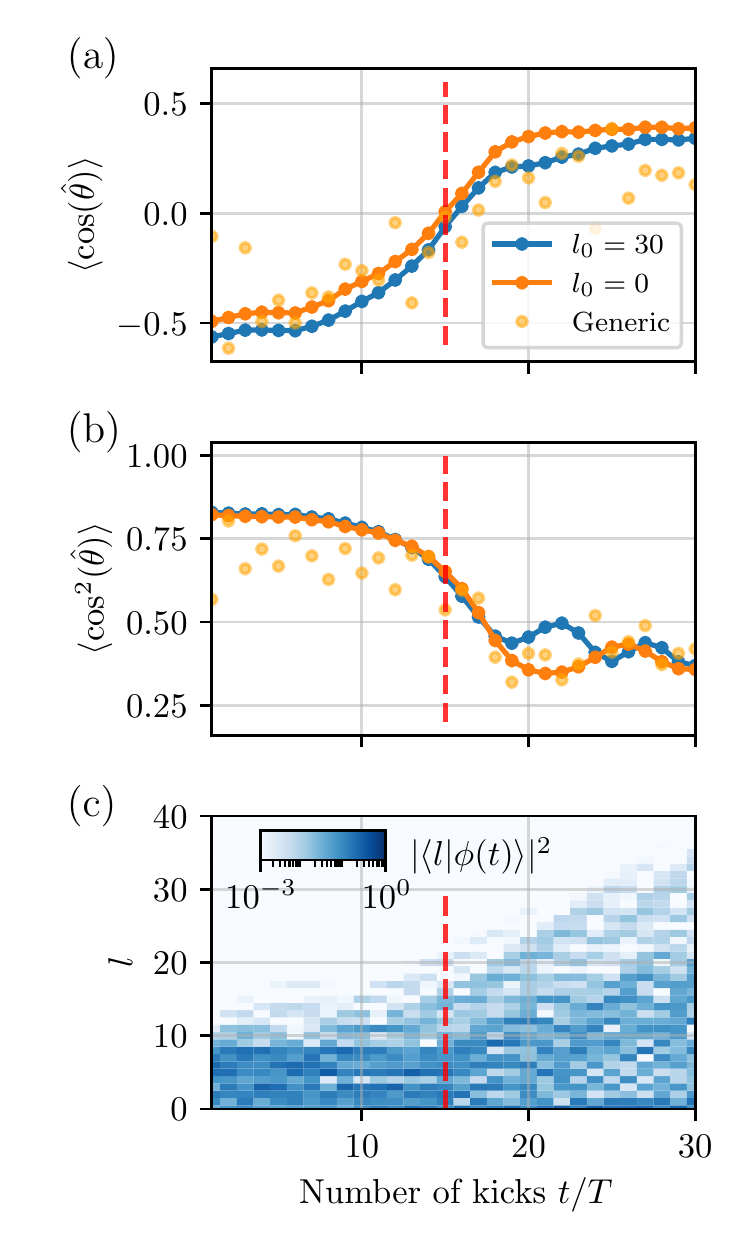}
    \caption{Time evolution of three different molecular states initially prepared in the third band of the spectrum peaked at $k=\pi$, see Fig.~\ref{fig:Fig2}(a), which are evolved through the Dirac cone at $t_c=15$, cf.\ Eq.~\eqref{eq:alpha}. To demonstrate that the phenomenon is generic we show a Gaussian state peaked at $l_0=30$ (orange), $l_0=0$ (blue) and a generic state prepared with one laser pulse from $l=0$ (dotted). As a function of the number of laser kicks are shown: (a) the orientation cosine, $\langle \cos(\hat{\theta)} \rangle$, (b) the alignment cosine, $\langle \cos^2(\hat{\theta)} \rangle$, and (c) the absolute values squared of the wavefunction components $\vert \langle l \vert \phi(t) \rangle \vert^2$ of the ``generic'' state. The dynamics changes drastically in the vicinity of the Dirac cone, marked by vertical dashed lines (red), see the text. The reason for this is a monopole at $k=\pi, \alpha=\alpha_c$ which changes the nature of the eigenstates.}\label{fig:quench}
\end{figure}

In Fig.~\ref{fig:quench} we show the results of this calculation for an initial Gaussian states with width $\Delta l=0.5$ and peak at $l_0=30$ and $l_0=0$ respectively. Furthermore, with yellow dots we show the results for a ``generic'' initial state created by a single laser pulse from $\phi = \delta_{l,0}$. {``Generic'' implies that we are choosing a state which can be created by one laser pulse starting from $l=0$. Evidently, to see a difference in the alignment/orientation signal, one needs to start from a state which is somewhat aligned or oriented~\footnote{We choose $\vert \phi \rangle = e^{-i \Delta t \hat{L}^2}e^{iV(P_1,P_2)}|l=0\rangle$ with $\Delta t\approx 0.911, P_1 \approx 6.67, P_2 \approx 2.67$. However, any other state with finite occupation in one band around $k=\pi$ can be used.}}. Shown are: (a) the orientation cosine, $\langle \cos(\hat{\theta}) \rangle \equiv \langle \phi(t)|\cos(\hat{\theta})|\phi(t) \rangle$, (b) the alignment cosine, $\langle \cos^2(\hat{\theta}) \rangle \equiv \langle \phi(t)|\cos^2(\hat{\theta})|\phi(t) \rangle$ and (c)~the absolute values squared of the wavefunction components of the ``generic'' state created by a pulse, $\vert \langle l \vert \phi(t) \rangle \vert^2$. 
These quantities, expressed through the populations of rotational states and molecular axis distributions in real space, are being routinely measured in gas-phase molecular experiments, e.g., using Coulomb explosion~\cite{ChatterleyPRL20} and Raman spectroscopy~\cite{FlossPRA18}, for (a,b) and (c), respectively.

In the initial time evolution, we observe a stagnant phase with little growth in energy and change in the alignment traces, which is due to the occupation within a specific band. Near the Dirac cones, however, the behavior changes, and we see a change of the orientation and alignment traces. After the cone, the generic behavior predicted at a quantum resonance, namely the linear spread of the wavefunction in angular momentum space is observed. This  can also be intuitively understood in terms of a quantum walk in the tight-binding description of the lattice. 
If we assume that the overlap of the initial wavepacket with the edge states is vanishing, we can expand its Fourier transform  in eigenstates with $\phi_k(t=0) = \sum_n \xi_n\psi_n(k,0)$, see Eq.~\eqref{eq:fourier}. The orientation and alignment traces change their behavior at the Dirac cones due to the change of the eigenstates. We note that for a ``generic'' wave packet which is not prepared exactly in the third band, this phenomenon is still visible, although not as pronounced, depending on the particular setup, see Fig~\ref{fig:quench}.
Finally we note that the example we are looking at with $\lesssim 30$ pulses is currently accessible in laboratory, as 24 kicks were used to observe dynamical localization with nitrogen molecules~\cite{bitter_experimental_2016}.

To summarize, we have demonstrated the possibility to observe topological {charges} in periodically kicked molecules. A key step connecting molecular rotational spectroscopy with single-particle models of condensed matter was to introduce quasimomentum, in addition to quasienergies of the Floquet representation. As opposed to topological states in  solid-state systems, which usually show up through some  macroscopic observables~\cite{bernevig_topological_2013,wang_topological_2017}, and topological conical intersections that can alter chemical reaction dynamics~\cite{ConicalBook, RyabinkinACR17}, the topological {charges} described here can be probed directly by imaging  molecular rotational states. In addition to new insights into the topological physics in molecular systems, this paves the way to {engineer} chemically relevant states of molecules using topological invariants. Unlike in many other topological systems, the position of the Dirac cones in molecules can be controlled directly by changing the parametrization of the laser strengths. Furthermore, laser pulses can be applied in a way selectively breaking the symmetries (e.g.\ time-reversal). The richness of internal degrees of freedom even for diatomic molecules~\cite{BrownRot} along with the possibilities to coherently control them~\cite{koch_quantum_2019} makes the extensions of the model to {higher dimensions possible. Using elliptically-polarized laser pulses the space of rotations for an asymmetric molecule becomes three dimensional (in the quantum numbers $l,m,k$) and a hopping in all the three dimensions can be induced~\cite{pabst_computational_2010}. Applying pulses which are in resonance with the rotational periods, multi-band and possibly non-abelian physics in more than one physical dimension could be probed. Lastly, previous works indicate that immersing rotating molecules into superfluid helium~\cite{lemeshko2016molecular} might generate an interacting topological system with non-abelian invariants~\cite{yakaboylu2017emergence}}. M.L.~acknowledges support by the European Research Council (ERC) Starting Grant No.~801770 (ANGULON). 
\bibliography{main}
\end{document}



\newcommand{\with}{\quad \text{with}\quad}
\newcommand{\where}{\quad \text{where}\quad}
\renewcommand{\and}{\quad \text{and}\quad}
\newcommand{\ie}{i.\,e. }
\newcommand{\Ie}{I.\,e. }
\newcommand{\eg}{e.\,g. }
\newcommand{\Eg}{E.\,g. }
\renewcommand{\vec}{\mathbf}
\renewcommand{\d}{\mathrm{d}}
\renewcommand{\t}[1]{\mathrm{#1}}
\newcommand{\im}{\mathrm{i}\mkern1mu}
\newcommand{\tr}{\mathrm{tr}}

\author{Volker Karle, Areg Ghazaryan, Mikhail Lemeshko}
\affiliation{Institute of Science and Technology Austria, Am Campus 1, 3400 Klosterneuburg, Austria}
\title{\papername}
\maketitle

\setcounter{equation}{0}
\setcounter{figure}{0}
\setcounter{table}{0}
\setcounter{page}{1}
\setcounter{section}{0}

\makeatletter
\renewcommand{\theequation}{S\arabic{equation}}
\renewcommand{\thefigure}{S\arabic{figure}}
\renewcommand{\thepage}{S\arabic{page}}

\newcommand{\rtext}[1]{\textcolor{red}{{#1}}}
\newcommand{\btext}[1]{\textcolor{blue}{{#1}}}
\newcommand{\gtext}[1]{\textcolor{orange}{{#1}}}
\newcommand{\oldtext}[1]{\textcolor{blue}{\sout{#1}}}
\section{Technical details}
\subsection{The sudden approximation for short half-cycle pulses}
\begin{SCfigure}
  \centering
  \includegraphics[width=0.58\textwidth]{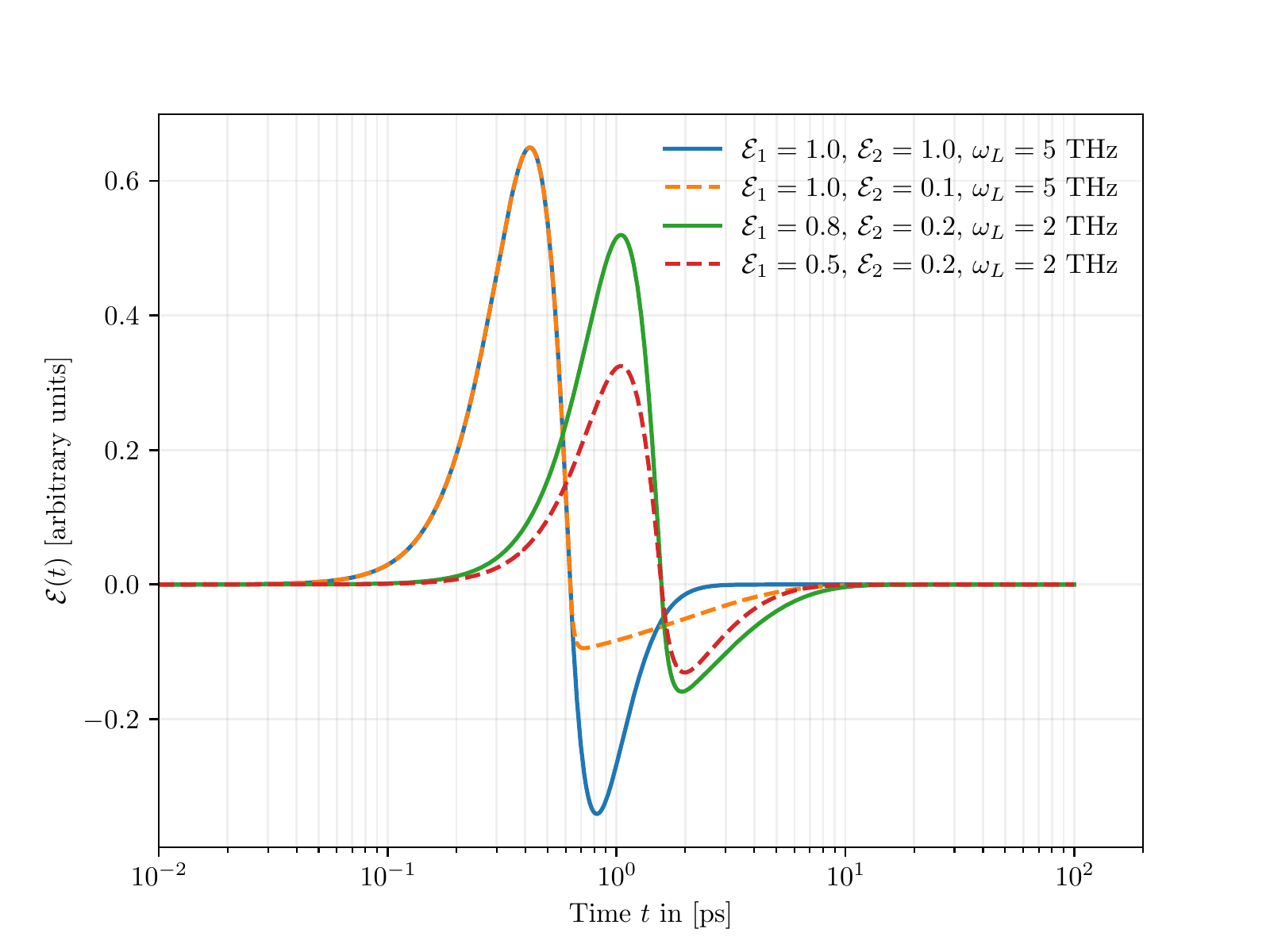}
  \caption{The pulse shape of~\eqref{eq:pulse_shape} taken from Ref.~\cite{barth2008nonadiabatic} for different parameters. The field amplitude $\mathcal{E}_1$ determines the height of the positive pulse, and $\mathcal{E}_2$ the depth of the negative tail. The pulse duration is determined by $t_p = \pi/\omega_L$ with the laser frequency $\omega_L$ and the decay time is given by the ratio $\mathcal{E}_2/\mathcal{E}_1$ (see main text);  a larger ratio leads to faster decay time, but larger negative peak, since the integral over $\mathcal{E}(t)$ is zero. For the sudden approximation to hold, this ratio needs to be small enough (\ie $\mathcal{E}_1 \gg \mathcal{E}_2$) to ensure an approximately positive Gaussian peak, otherwise the potential will vary considerably. Further, it is important that the pulse duration $t_p \ll \tau_B$ to ensure a separation of the timescales of the laser and the rotation.}
    \label{fig:pulse_shape}
\end{SCfigure}

\begin{figure}[htpb]
    \centering
    \includegraphics[width=1.05\textwidth]{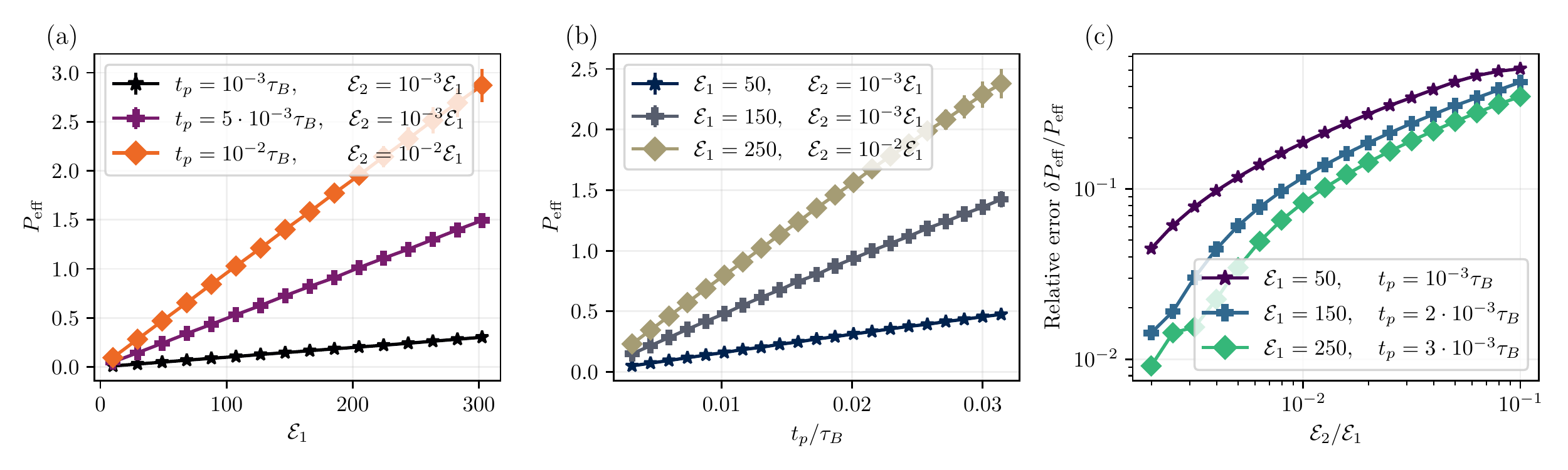}
    \caption{In (a) and (b) we show the dependence of the effective kicking strength $P_\mathrm{eff}$ on the half-cycle field amplitudes $\mathcal{E}_1,\mathcal{E}_1$ and the pulse duration $t_p$ (in units of the rotational period $\tau_B=\pi/B$), see~\eqref{eq:pulse_shape} for the parametrization of the half-cycle pulse. The effective kicking strength is calculated using the method in the text with $P_\mathrm{eff}=\langle p_l \rangle$ with the error $\delta P^2_\mathrm{eff} =\langle p_l^2\rangle - {\langle p_l\rangle}^2$ with a summation up to $l=50$ and a maximum $l_\mathrm{max}=150$. We observe that for the parameters shown here the errors are relatively small and the agreement with the sudden approximation is therefore very good. In (a), for increasing $\mathcal{E}_1$, the errors become larger in agreement with the expectation for large field amplitudes the exact time evolution becomes important. The same reasoning applies to (b), which shows that for vanishing pulse duration, the sudden approximation becomes exact. In (c) we show how the relative error $\Delta P_\mathrm{eff}/P_\mathrm{eff}$ scales with the ratio $\mathcal{E}_2/\mathcal{E}_1$. We see that it falls off algebraically in the limit $\mathcal{E}_2/\mathcal{E}_1 \rightarrow 0$. This shows that in this limit the full time-evolution converges to the sudden approximation. The reason is that the decay time is controlled by exactly this ratio, and for decreasing ratios the decay time goes to infinity. The reasoning is that for decay times larger than the pulse duration $t_p$, the sudden limit is valid. A similar convergence can be observed for the limits $t_p/\tau_B \rightarrow 0$ and the trivial limit $\mathcal{E}_{1,2}\rightarrow 0$. We note that in order to observe the convergence of the $\mathcal{E}_2/\mathcal{E}_1 \rightarrow 0$ limit, the pulse duration needs to be small enough, i.e. $t_p \ll \tau_B$. Further, the limitations of the approximation which were mentioned in the main text remain, \ie~the approximation only holds up to $l$ with $\omega_\mathrm{rot}(l) \ll \omega_L$.}
    \label{fig:Peff}
\end{figure}

In this section we  justify the use of the sudden approximation for short half-cycle pulses. For intense laser pulses that are much shorter than the rotational timescale, it is possible to model the laser potential as a delta-pulse~\cite{dion2001orienting,seideman1999revival,dion2002optimal}. Separating the time scale of the rotation of the molecule $\tau_B = \pi/B$ from the fast oscillations of a single laser pulse simplifies the time-dependent Schrödinger equation. In this work we make use of two kinds of intense off-resonant laser pulses: (i) short multi-cycle infrared pulses that couple to the molecular polarizability and (ii)
half-cycle terahertz pulses that couple to the permanent dipole moment of the molecule. The half-cycle pulse is typically treated as a strong unipolar pulse $\propto P_\text{eff} \cos(\theta)$, since one intense peak is followed by a long, negative and weak tail~\cite{dion_laser-induced_1999,mirahmadi2018dynamics}, see Fig.~\ref{fig:pulse_shape}. However, while the validity of the sudden approximation for infrared pulse trains (i) is well studied in theory and experiment~\cite{bitter_experimental_2016,bitter_control_2017,lin_visualizing_2015}, it remains to be shown for half-cycle pulses (ii). Theoretically, the question is whether the time-evolution operator that includes the full time-evolution of the laser intensity can be approximated by a simplified propagator. For a rigid rotor driven by a linearly polarized laser pulse with field amplitude $\mathcal{E}(t)$, the Schrödinger equation reads~\cite{dion_laser-induced_1999,dion2001orienting}:
\begin{equation}
  \im \partial_t |\psi(t)\rangle = \hat{H}(t)|\psi(t)\rangle \with \hat{H}(t) = B \hat{\mathbf{L}}^2 - \mu_0 \mathcal{E}(t) \cos(\hat \theta),
\end{equation}
with rotational constant $B$, angular momentum operator $\hat{\mathbf{L}}^2$ and permanent dipole moment $\mu_0$. In many works, the field amplitude is assumed to take the form of a Gaussian. However, for realistic experiments with molecular beams, the integral of field has to vanish in the far-field regime~\cite{rauch2006time,arkhipov2022half,barth2008nonadiabatic}, \ie $\int^{\infty}_{-\infty}\mathcal{E}(t) \d t = 0$. In that case, it is not trivial to show that such a pulse can be approximated by an instantaneous pulse given the long negative tail of the field. It turns out that the validity of the approximation strongly depends on the initial state, the energy transferred to the molecule and the typical timescales (i.e. the pulse duration and the decay time) of the laser. To elaborate further on this matter, we choose a parametrization of the field amplitude with zero integration measure~\cite{barth2008nonadiabatic},
\begin{equation}
   \mathcal{E}(t) = 
\begin{cases}
0 & (t \leq 0)\\
\mathcal{E}_1 \cos^2(\omega_L (t-t_p)/2)\sin(\omega_L(t-t_p)) & (0 \leq t < t_p)\\
\mathcal{E}_2 \left(1 - e^{-(t-t_p)/\tau_1} \right) e^{-(t-t_p)/\tau_2} & (t \geq t_p),
\end{cases}
\label{eq:pulse_shape}
\end{equation}
with electric field amplitudes $\mathcal{E}_1, \mathcal{E}_2>0$ (dimensionless in units of $B\equiv \mu_0 \equiv \hbar  \equiv 1$), the laser frequency $\omega_L$, the pulse duration of the first part of the laser pulse $t_p$, the switch-on and switch-off times $\tau_1, \tau_2$. Here all quantities are dimensionless, but we easily can convert the dimensionless $\mathcal{E}$ to physical units by noting that $B=\hbar=1$ rescales time as $\tilde{t} = B \cdot t/ \hbar$. Then, the time-evolution of the potential reads $\exp[-i \hat{V} \tilde{t}/B]$, hence the conversion factor of a field $E$ with units of kV/cm is 
\begin{equation}
  E \, [\mathrm{kV/cm}]  =  59.55 \,[\mathrm{kV}\cdot\mathrm{Debye}]\cdot B[\mathrm{cm}^{-1}]/\mu_0[\mathrm{Debye}]\cdot \mathcal{E}[\mathrm{dimensionless}]
\end{equation}
For the OCS molecule frequently used in short-pulse alignment experiments~\cite{ChatterleyPRL20}  ($B = 0.203$~cm$^{-1}$, $\mu_0 = 0.715$~D), this is for example around $\approx 17~\mathrm{kV}/\mathrm{cm}$. That implies that in order to get the physical units, one needs to multiply the dimensionless $\mathcal{E}$  (e.g.\ from the $x$-axis of Fig.~\ref{fig:Peff}(a)) by this factor.

As far as laser frequencies are concerned, the field has to be far-off-resonant with respect to all rotational and vibrational  transitions in the molecule, and have a much lower frequency as compared to the electronic excitations. In this case, at every moment of time during the pulse, the electrons in the molecule  instantaneously adjust  to the laser field  and the treatment  can be reduced to the effective interaction between the laser's electric field and the dipole moment (and, in the next order, dipole polarizability) of the molecule.  In order to reach the short-pulse regime,  $t_p \approx 1 \% \tau_B$, a terahertz laser with $\omega_L \approx 4~\mathrm{Thz}$ can be used. The condition that the electric field is smooth at $t=t_p$ further leads to $\tau_1 = \frac{\mathcal{E}_2}{\omega_L \mathcal{E}_1}$ and the zero integration measure leads to $\tau_2 = (2\omega_L^2\tau_1)^{-1} + \sqrt{(2\omega_L \tau_1)^{-2} + (\omega_L)^{-2}}$ (for more details see Ref.~\cite{barth2008nonadiabatic}). Clearly, the integration measure of the positive peak scales as $P_\mathrm{eff}\propto \mathcal{E}_1 \cdot t_p$ (the negative tail compensates for exactly this value). The decay time is determined by $\tau_2$.

\begin{SCfigure}
  \centering
  \caption{Time-evolution of a rotational wavepacket with $\langle l=0 | \psi(t=0) \rangle = 1$, driven by a half-cycle pulse train, calculated with (a) the exact time-evolution operator $U(t)$ and (b) the-evolution operator $U^\mathrm{sudd}(t)$ of the sudden approximation. The parameters of the half-cycle pulse are $\epsilon_1 = 30, \epsilon_2=2, \tau_p\cdot \pi = 0.3 \tau_B$ and $T=\tau_B/3$. The effective kicking strength (\ie~the number of transferred $\hbar$ per kick) is estimated (optimizing the overlap) as $P_\mathrm{eff}\approx2.879$. Since the pulse duration is comparatively long, one has to correct the effective pulse position for the sudden approximation to $t_{p,\mathrm{eff}}\cdot \pi \approx 0.222 \tau_B $ (\ie the operator is $U=e^{-\im(t-t_p)\hat{H}_0}e^{+i\hat{V}}e^{-\im t_p \hat{H}_0}$). We observe that at this level of accuracy it is impossible to see a difference between the two different time evolutions. For initial states at large $l$, it is possible distinguish the two operators. However, for realistic experimental setups it is sufficient to understand the qualitative behavior of the sudden approximation.}\includegraphics[width=0.6\textwidth]{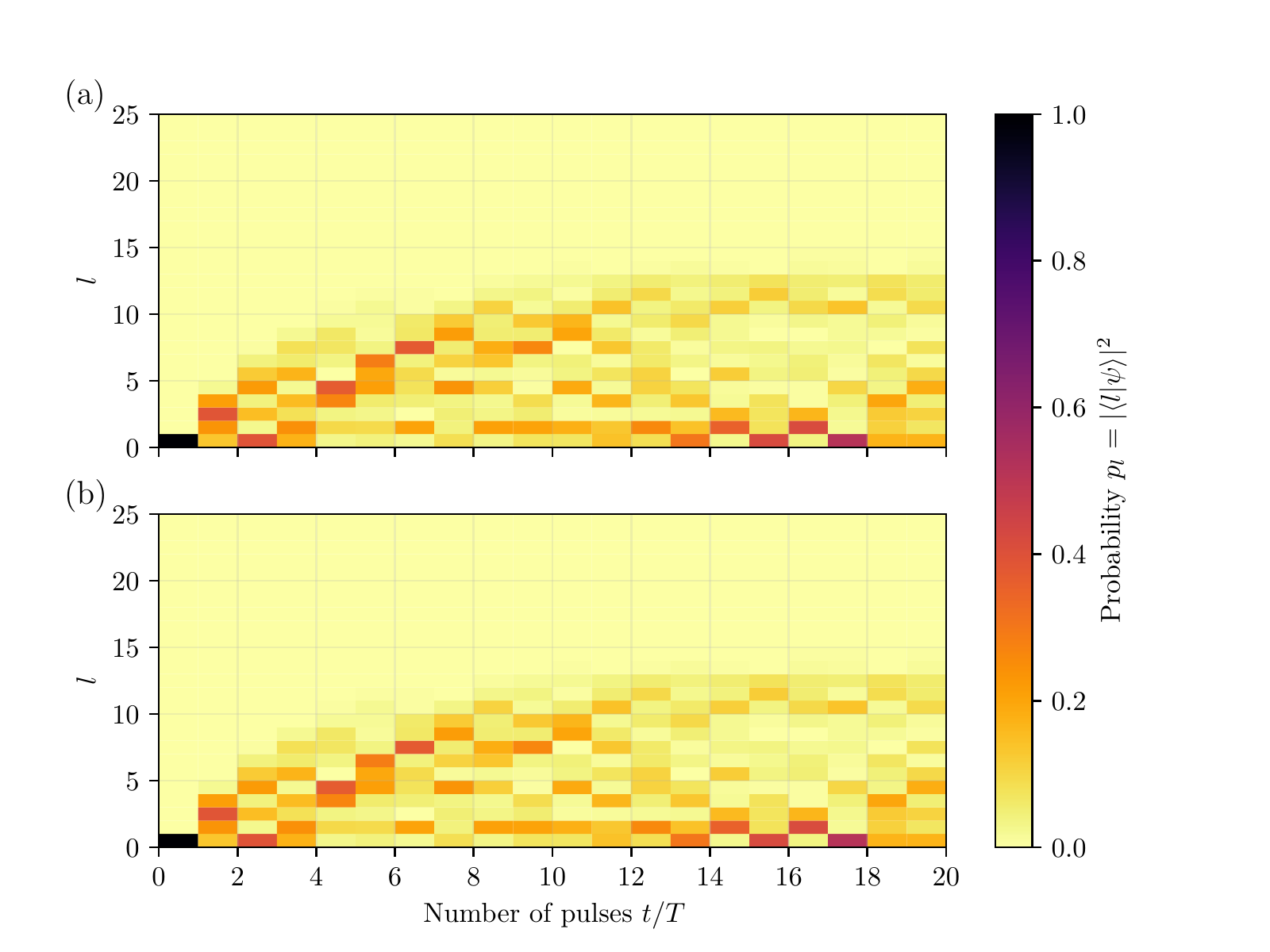}
    \label{fig:wave_function}
\end{SCfigure}

In order to study the validity of the sudden approximation, we numerically integrated the full differential equation of the time-evolution operator $\hat{U}(t)$ with  $\im \partial_t \hat{U}(t) = \hat{H}(t)\hat{U}(t)$ for a reasonable cut-off $l < l_\mathrm{max}$ and various parameters. Generally speaking, each angular momentum eigenstate oscillates with frequency $\omega_\mathrm{rot}(l)=\pi [l(l+1)]$ (in units of rotational times $\tau_B$), which leads to a natural cut-off scale; the approximation can only succeed for states with $\langle l|\psi \rangle \approx 0$ for $l$ with $\omega_L < \omega_\mathrm{rot}(l)$. The eigenstates $l$ with $\omega_L < \omega_\mathrm{rot}(l)$ oscillate with a frequency equal or higher than the laser frequency and a separation of timescales is not possible. For the sudden approximation with $\hat{U}^\mathrm{sudd}(t) \approx e^{-\im(t-t_p)\hat{H}_0}e^{i \hat{V}_\mathrm{sudd}}e^{-\im t_p \hat{H}_0}$ where $\hat{V}_\mathrm{sudd} = P \cos(\hat{\theta})$, only the off-diagonal matrix element $\langle l \pm 1|\hat{V}_\mathrm{sudd}|l \rangle$ are non-zero (c.f. Eq.~\eqref{eq:S1}). 
Whether or not the full time evolution can be approximated by that expression depends therefore on the question to what extend the matrix elements of  $V_\mathrm{eff} = -\im \log[e^{+\im(t-t_p)\hat{H}_0}\hat{U}(t)e^{+\im t_p \hat{H}_0}]$ resemble the ones of $V_\mathrm{sudd}$ for $t \gg t_p$ and $l<l_\mathrm{max}$. In Fig.~\ref{fig:wave_function} we calculate the time-evolution of a rotational wavepacket with (a) the full propagator and (b) the sudden approximation propagator. In that case, the approximation works very well and the two trajectories are practically the same.

To analyze in detail in what regime the sudden approximation works well is out of the scope of this work, however, we would like to shortly discuss the convergence of the full propagator to the propagator of the sudden approximation in certain limits. To this end, we compare the matrix element of the full propagator $v_l = \langle l + 1 | V_\mathrm{eff}| l\rangle$ with the exact term $w_l \equiv \langle l + 1 | \cos(\hat{\theta}) | l \rangle$ and calculate the ratio $p_l = v_l/ w_l $. For realistic scenarios, we care about the low angular momentum regime $l < l_\mathrm{mean}\ll l_\mathrm{max}$ and estimate $P_\mathrm{eff}$ with the average $P_\mathrm{eff}= \langle{p_l}\rangle$ and the error with $\delta P_\mathrm{eff} =\sqrt{\langle p_l^2\rangle - {\langle p_l\rangle}^2}$, where the averages are defined as sums up to some $l_\mathrm{mean}$, depending on the experimental setup and the number of kicks. In the case when the matrix-elements of the full propagator converges to the approximation, we expect $\delta P_\mathrm{eff}/P_\mathrm{eff} \rightarrow 0$. Numerically, this is what we  numerically observe for the limits $\mathcal{E}_1 \rightarrow 0$, $\mathcal{E}_2/\mathcal{E}_1 \rightarrow 0$, and $t_p \rightarrow 0$, see Fig.~\ref{fig:Peff}. These results indicate that the sudden approximation is a valid assumption in a number of cases. Specifically, the ratio $\mathcal{E}_2/\mathcal{E}_1$ strongly depends on the specific mechanism of the pulse generation and the switch-off time. On top of that, the characteristics of the laser with frequency $\omega_L$ already govern for which electric field amplitudes the sudden approximation can be used. However, the existence of Dirac cones in the Floquet spectrum does not rely on the strength of laser pulses alone, but on the ratio $P_1/P_2$, see Fig.~\ref{fig:gapclosings} and the corresponding text. Therefore, even for small field amplitudes it is possible to observe Dirac cones, given that the IR pulse strength is of the same order of magnitude as the Terahertz pulse in terms of transferred $\hbar$ per pulse (which does not depend on $\mathcal{E}_{1,2}$ alone, but also on $t_p$).

\subsection{Fourier transforms of angular momentum lattices}
\begin{figure}[t]
        \centering
        \includegraphics[width=0.9\textwidth]{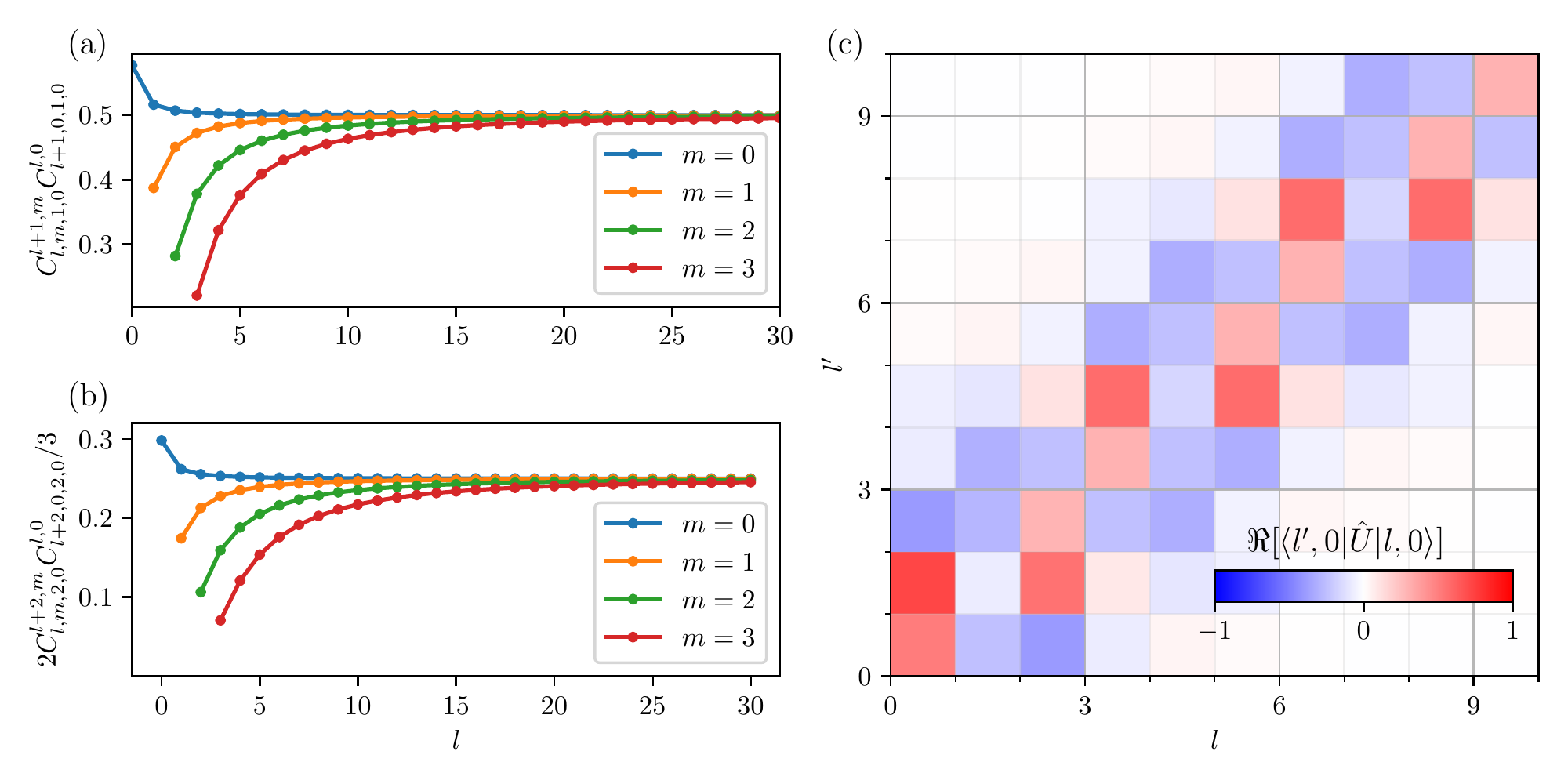}
        \caption{Convergence of the hopping elements for $l,l'\gg0$ to a constant value. In (a) we show the element which is relevant for the $\cos(\theta)$ potential and in (b) for the $\cos^2(\theta)$. In (c) we show the real part of the \textit{effective hopping elements} $\langle l' | U | l\rangle$ where the time-translation operator is $\hat{U} = e^{- \pi \im \hat{\mathbf{L}}^2/ N}e^{\im \hat{V}(\hat{\theta})}$, with $N=3, P=2.5$ and $\alpha=\pi/3$ for the potential $V(\hat{\theta}) = P \cdot \left (\cos^2(\alpha)\cos^2(\hat{\theta}) + \sin^2(\alpha)\cos(\hat{\theta})\right)$ as in the rest of the letter. We marked the unit-cell size $N=3$ by gray lines. Note that the hopping elements are inherently complex, as the driven molecule is a non-hermitian system.}
        \label{fig:matrixelements}
    \end{figure}

Here we show that the Fourier transform of the angular momentum lattice provides a viable tool to analyze the Floquet spectra at quantum resonances of a periodically driven molecule. As elaborated in the main text, at a quantum resonance, the time-translational operator takes the form
    $\hat{U} = e^{- \pi \im \hat{\mathbf{L}}^2/ N}e^{\im \hat{V}(P_1,P_2)}$. For odd $N$, the rotational part is periodic with $N$, while for even $N$ the periodicity is $2N$, which becomes evident when looking in the $l$-basis with $\langle l | e^{- \pi \im \hat{\mathbf{L}}^2/ N} | l' \rangle  = e^{-i\pi l(l+1)/N}\delta_{ll'}$. Our approach works for all laser potentials which converge to a constant value, \ie $V_{l,l'} \approx V_{l+1,l'+1}$, which is in particular true for the combination of a ``regular'' multi-cycle laser pulse (the $\cos^2(\hat{\theta})$ term) and a half-cycle laser pulse (the $\cos(\hat \theta)$ term), which can be proven using the Edmonds asymptotic formula for the $3j$-symbols~\cite{flude_edmonds_1998}. The asymptotic expressions take the simple form \ie 
    \begin{align}
      \langle l'm'|\cos(\theta)|lm\rangle &= -\delta_{mm'}C^{l'm}_{lm10}C^{l0}_{l'010} \xrightarrow[]{\text{for $l,l'\gg 0$}}\delta_{mm'}\left(\delta_{l,l'+ 1} + \delta_{l,l'- 1}\right)/2 \label{eq:S1}\\
      \langle l'm'|\cos^2(\theta)|lm\rangle &= \delta_{mm'}\left (\tfrac{2}{3}C^{l'm}_{lm20}C^{l0}_{l'020} + \tfrac{1}{3}\delta_{ll'} \right)  
      \xrightarrow{\text{for $l,l'\gg 0$}}\delta_{mm'} \left(\delta_{l,l'} + (\delta_{l,l'+2} +\delta_{l,l'-2})/2\right) /2
      \label{eq:S2}
    \end{align}
    with the usual Clebsch–Gordan coefficients $C^{LM}_{l'm'lm}=\langle l',l,;m',m | l',l; L,M\rangle$. To prove this asymptotic, one can use the Edmonds asymptotic formula for 3j-symbols~\cite{flude_edmonds_1998}  
    \begin{equation}
      {\begin{pmatrix}l_{1}&l_{2}&l_{3}\\m_{1}&m_{2}&m_{3}\end{pmatrix}}\xrightarrow[]{\text{for $l_{2},l_{3}\gg l_{1}$}}(-1)^{l_{3}+m_{3}}{\frac {d_{m_{1},l_{3}-l_{2}}^{l_{1}}(\theta )}{\sqrt {l_{2}+l_{3}+1}}} \with \cos(\theta) = \frac{m_2-m_3}{l_2+l_3+1},
    \end{equation}
    where $d^l_{m,m^\prime}(\theta)$ is the Wigner function and 3j-symbols are related to the Clebsch-Gordan coefficients~\cite{varshalovich_quantum_1988}
    \begin{equation}
\langle j_{1}\,m_{1}\,j_{2}\,m_{2}|J\,M\rangle =(-1)^{-j_{1}+j_{2}-M}{\sqrt {2J+1}}{\begin{pmatrix}j_{1}&j_{2}&J\\m_{1}&m_{2}&-M\end{pmatrix}}.
    \end{equation}
    For Eq.~\eqref{eq:S1} this leads to 
    \begin{equation}
        C^{l'm}_{lm10}C^{l0}_{l'010} = (-1)^{-m}\sqrt{(2l'+1)(2l+1)}\begin{pmatrix}
        1 & l' & l \\
        0 & m & m 
        \end{pmatrix} 
\begin{pmatrix}
        1 & l & l' \\
        0 & 0 & 0 
        \end{pmatrix} 
        \xrightarrow[]{\text{for $l,l'\gg 0$}}(-1)^{l+l'} d^1_{0,l-l'}(\sfrac{\pi}{2})d^1_{0,l'-l}(\sfrac{\pi}{2})
    \end{equation}
    where we used that $\theta = \pi/2$ for both terms.
    Since the Clebsch-Gordan coefficients allow only $l-l'=\pm 1$ and $d^1_{0,\pm 1}(\sfrac{\pi}{2})d^1_{0,\mp 1}(\sfrac{\pi}{2})=\sfrac{1}{2}$ this leads to the desired relation. For Eq.~\eqref{eq:S2} this derivation proceeds analogously, with the difference that  $d^2_{0,\pm 2}(\sfrac{\pi}{2})d^2_{0,\mp 2}(\sfrac{\pi}{2})=\sfrac{3}{8}$ and $d^2_{0,\pm 2}(\sfrac{\pi}{2})d^2_{0,\mp 2}(\sfrac{\pi}{2})=\sfrac{3}{8}$ and $d^2_{0,0}(\sfrac{\pi}{2})d^2_{0,0}(\sfrac{\pi}{2})=\sfrac{1}{4}$, which in total leads to the desired result. For the projection of angular momentum $m=0$ these coefficients converge fast, see Fig.~\ref{fig:matrixelements}, and the overall behavior of the system is very well described by an approximation where we assume a constant value. In this work we only consider $m=0$, but for other $m$ the approximations work out as well when considering larger $l$.
    The $N$-periodicity allows us to define a Fourier transform which approximates the periodic behavior for $l,l' \gg 0$. Any operator $\hat{A}$ with that periodicity, and the effective Hamiltonian in particular, follows
   \begin{equation}
      \langle l' | \hat{A}|l \rangle \approx \langle l'+N | \hat{A}|l+N \rangle 
      \label{eq:period}
   \end{equation}
for $l,l'\gg 0$. Let us parametrize each $l',l$ by $l=n\cdot N + i, l'=n'\cdot N + j$ with $n,n' \in \mathbb{N}_0,i,j \in \{0,1,\dots,N\}$. Then, 
\begin{equation}
  A_{l'l} = A_{i,j}(n',n)=A_{i,j}(n'-n)=A_{i,j}(\Delta n)  
\end{equation}
\ie $A$ only depends on the off-diagonal index $\Delta n$ because of the periodicity. The Fourier transform from $l$- to $k$-space then becomes (for a dimensionless momentum variable $k \in [0,2\pi]$)
\begin{equation}
\begin{aligned}
\label{eq:Aeff}
    \underbrace{A_{ij}(k',k)}_{N\text{x}N \text{Matrix}} &= \sum_{n',n} e^{-\im ( n'\cdot k' - n \cdot k)}  A_{ij}(n',n) = \sum_{\Delta n, \bar{n}}e^{-\im (\bar{n}(k'-k)/2 + \Delta n (k'+k)/2)} A_{ij}(\Delta n) 
 = \delta_{kk'}\sum_{\Delta n} e^{-\im \Delta n \cdot k}  A_{ij}(\Delta n),  \\
\text{i.e.} \quad A_{ij}(k) &=\sum_{\Delta n} e^{-\im \Delta n \cdot k} A_{ij}(\Delta n) \quad \text{and correspondingly}\quad A_{ij}(\Delta n) =\frac{1}{2\pi} \int_0^{2\pi} A_{ij}(k)\d k.
\end{aligned}
\end{equation}
where we used $\bar{n}=n'+n, \, \Delta n = n'-n$ and $A_{ij}(n',n)=A_{ij}(\Delta n)$, which shows that operator $\hat{A}$ conserves momentum $k$. Since the Fourier transform is only well defined for an infinite system, for a finite system this implies practically an infrared cutoff in momentum which is defined by the system size. When calculating the Fourier transform of an operator, one needs to choose a unit-cell in the middle of the lattice to avoid edge effects. Practically, we choose $A_{ij}(\Delta n) = A_{N \cdot n_0+i,N\cdot (n_0 + \Delta n) +j}$ with a large enough $n_0$ to ensure convergence. For an infinite and perfectly translationally invariant system, we would sum over all $\Delta n \in \mathbb{Z}$. However, here we need to choose $-n_0 \ll \Delta n \ll n_0$. For small to moderate $P$, only few off-diagonals suffice to achieve high agreement with the ``real-space'' diagonalization. A state transforms with
\begin{equation}
   f_i(k) = \sum_{\Delta n} e^{-\im \Delta n \cdot k} f(i + N(n_0 + \Delta n)), \quad f(i+N(n_0 +\Delta n)) =\frac{1}{2\pi} \int_0^{\infty}f_i(k)e^{+\im \Delta n \cdot k}\d k.
   \label{eq:fourier}
\end{equation}
The fourier transform of the operators at $l',l \gg 0$ in~\eqref{eq:S1} and~\eqref{eq:S2} is given by 
\begin{equation}
  (\cos(\hat{\theta}))_{ij}(k) = \frac{1}{2}\underbrace{\begin{pmatrix}
0 & 1 & & &  & & \dots& & e^{-ik}\\
1 & 0 & 1 &  & & & & & \\
 & 1 & 0 &  & & & & & \\
\vdots  &   &   & \ddots & & & & & \vdots \\
  &  &  &  & & & & 1& \\
 & &  & & & &1 & 0 &1\\
e^{+ik} && \dots & & & & &1 & 0\\
\end{pmatrix}}_{N \times N\text{ Matrix}} 
\end{equation}
\begin{equation}
  \text{and}\quad (\cos^2(\hat{\theta}))_{ij}(k) = \frac{1}{2}\underbrace{\begin{pmatrix}
1 & 0 & \sfrac{1}{2}& &  & & \dots& e^{-ik}/2& 0\\
0 & 1 & 0 & \sfrac{1}{2} & & & &\dots & e^{-ik}/2\\
\sfrac{1}{2} & 0 & 1 & 0 & \sfrac{1}{2}& & & & \\
& \sfrac{1}{2} & 0 & 1 & 0 & & & & \\
\vdots  &   &   & \ddots & & & & & \vdots \\
  &  &  &  & & &1 & 0&\sfrac{1}{2} \\
 e^{+ik}/2& &  & & & &0 & 1 &0\\
0 & e^{+ik}/2 & \dots & & & &\sfrac{1}{2} &0 & 1\\
\end{pmatrix}.}_{N\times N\text{ Matrix}} 
\end{equation}
To calculate $e^{\im \hat{V}}$ it is sufficient to diagonalize $V_{i,j}(k)$ and express the exponential in terms of its eigenstates $v_n(k)$ and eigenvalues $\lambda_n(k)$, \ie 
   $e^{\im V(k)} = \sum_{n} e^{\im \lambda_n(k)}\left (v_n(k) \otimes v^\dagger_n(k)\right ).$
\subsection{Symmetries}
\begin{figure}[t]
    \centering
    \includegraphics[width=0.9\textwidth]{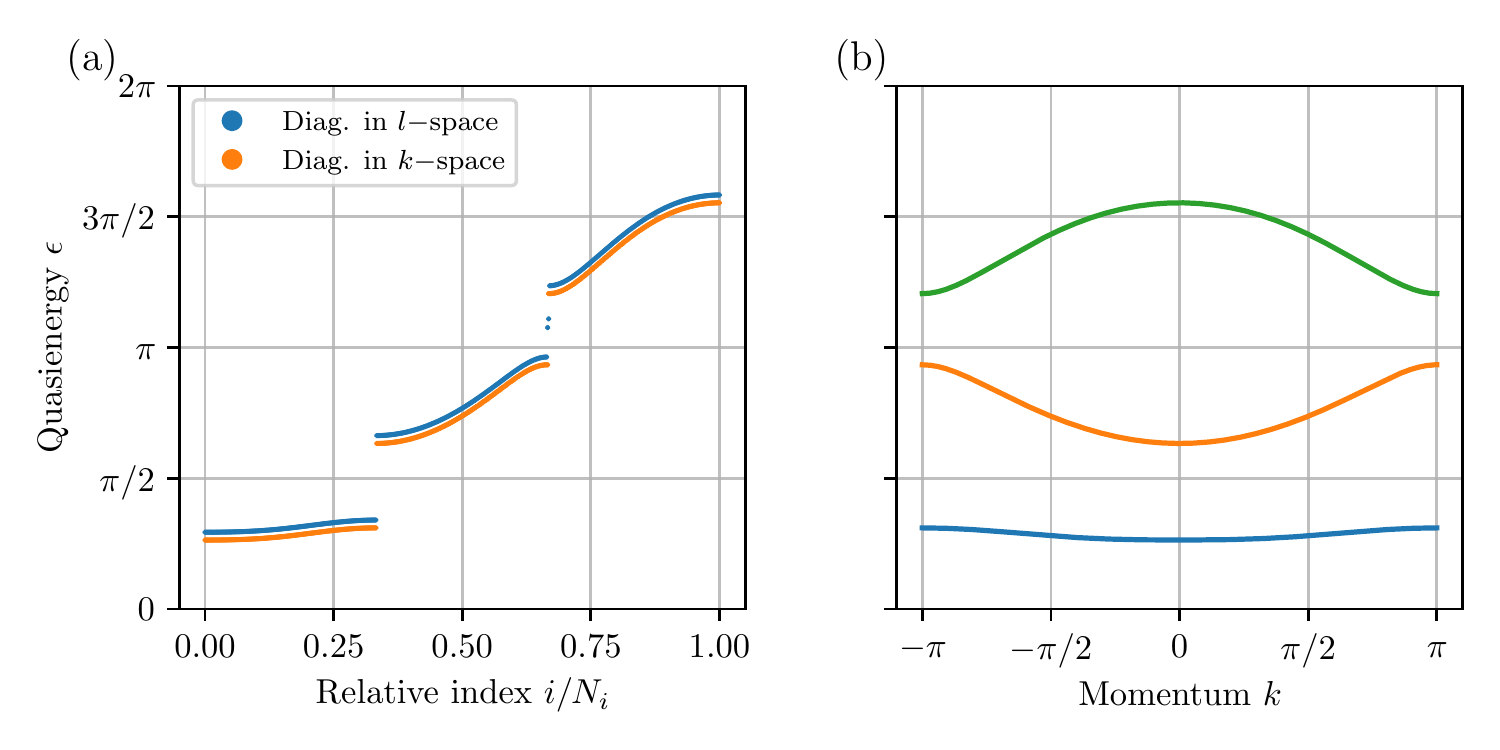}
    \caption{Comparison of  two diagonalization methods for $P_1=1.5,P_2=1.8$. The left figure shows the quasienergies sorted by their relative index, and in the right figure depending on their momentum $k$. The agreement of the two diagonalization methods are high, implying that the bulk behavior dominates the spectrum. In the left figure the diagonalization in $l-$space gives rise to two in-gap states, which are the edge modes mentioned in the main text. They cannot be obtained using the momentum space diagonalization for obvious reasons. Also visible in the spectrum obtained in $k-$space is the reflection symmetry, which guarantees $\epsilon_k = \epsilon_{-k}$.}
    \label{fig:fourier1}
\end{figure}
The rotational phases, $\phi_{l}=e^{-\pi l(l+1)/N}$, give rise to reflection symmetry for all $N$ with translationally-invariant potentials (\ie in our case for $l',l \gg 0$). For an odd $N$, the reflection center of the first unit cell is $n_c = (N+1)/2 \in \mathbb{N}$. Then, for a $0 \le j < n_c$ we have $\phi_{n_c + j} = \phi_{n_c - j}$ (for even $N$, the unit-cell is $2N$ and $n_c = N$ with $\phi_{n_c + j+1} = \phi_{n_c - j}$). If we limit ourselves to a finite Hilbert space with maximum $l_\mathrm{max}$, we can write the symmetry in the $l$-basis as
\begin{equation}
   \mathcal{R} = \underbrace{\begin{pmatrix} 
    & &\dots &0 & 1\\ 
    &  & & 1 & 0\\ 
  \vdots  &  & \iddots& & \vdots \\ 
  0 & 1& &&  \\ 
  1 & 0& \dots & \end{pmatrix}}_{(l_\mathrm{max}+1)\times(l_\mathrm{max}+1)\text{ Matrix}}
\end{equation}
with $\mathcal{R}^\dagger \mathcal{R} = \mathcal{R}^2 = \mathbb{I}$. It commutes with the time-translation operator, $[U, \mathcal{R}] = 0$. In terms of the effective Hamiltonian, $\hat{H}= \im \log \hat{U}$, this symmetry reads in $k$-space $\mathcal{R}H_k = H_{-k}\mathcal{R}$ with
\begin{equation}
   \mathcal{R} = \underbrace{\begin{pmatrix} 
    & &\dots &0 & 1\\ 
    &  & & 1 & 0\\ 
  \vdots  &  & \iddots& & \vdots \\ 
  0 & 1& &&  \\ 
  1 & 0& \dots & \end{pmatrix}}_{N \times N \text{ Matrix}} \mathcal{P} \quad {\text{and the parity operator with}} \quad \mathcal{P}\phi_k = \phi_{-k}.
\end{equation}
For the one-kick system, shifting the position of the kick does not change the model, \ie
\begin{equation}
    U(\beta) = e^{-\im \hat{\mathbf{L}}^2\pi(1-\beta)/N }e^{\im V_1}e^{-\im \hat{\mathbf{L}}^2\pi\beta/N } \quad \sim \quad U = e^{-\im \hat{\mathbf{L}}^2\pi/N }e^{\im V_1}
\end{equation}
where the $\sim$ implies that the two operators have the same spectrum up to a constant for any $0 \le \beta\le 1$.  However, as we have seen above, a finite $\beta$ leads to different $N-$periodicity. Let us write $\beta= \frac{a}{b}$ with $a,b \in \mathbb{N},a<b$. Then, for incommensurable $a,b\cdot N$, the periodicity of the operator will be $b \cdot N$ (correspondingly, for commensurable $a$ and $N \cdot b$ it will be the lowest common denominator). The equivalence of the spectra then leads to $bN$ bands in place of $N$ bands which compensates for the larger unit-cell. This is a well-known phenomenon commonly known as ``band folding'' in solid-state physics.\\\\
\begin{figure}[t]
    \centering
    \includegraphics[width=0.82\textwidth]{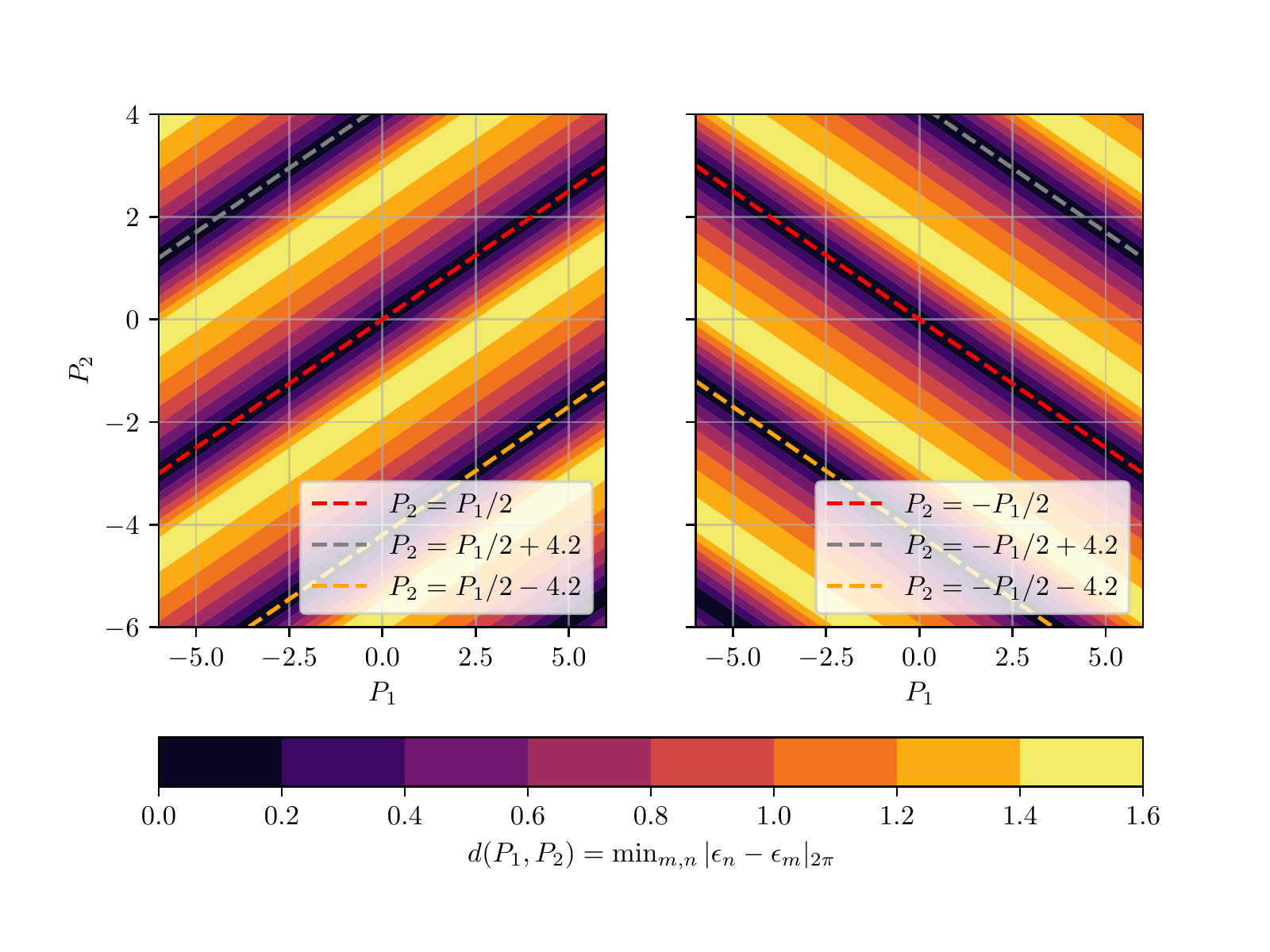}
    \caption{Smallest quasienergy differences (\ie smallest band gap) for $k=\pi$ (left figure) and $k=0$ (right figure). For small to intermediate couplings the only way to achieve gap closings is with $P_1 \neq 0, P_2 \neq 0$.}
    \label{fig:gapclosings}
\end{figure}
In addition to that, we find a time-reflection symmetry present for all models with one kick in the Floquet operator (where we use that our laser potentials follow $V^*=V = V^T)$. More specifically, the second kick (for instance, $U_\text{2-kick} = e^{-\im \hat{\mathbf{L}}^2\pi/N}e^{\im V_1}e^{-\im \hat{\mathbf{L}}^2\pi/N}e^{\im V_2}$ with $V_1 \neq V_2$ ) would break the time-reflection invariance. The time-reflection invariance is proportional to unity times conjugation when the pulse is at the center, \ie~$\beta= \sfrac{1}{2}$. In that gauge, we find
\begin{equation}
   \langle l' |U^*(\sfrac{1}{2})U(\sfrac{1}{2})|l \rangle = \sum_{l''}\langle l'|e^{+\im \pi l'(l'+1)/2N}e^{-\im V}|l''\rangle e^{+\im \pi l''(l''+1)/2N}e^{-\im \pi l''(l''+1)/2N}\langle l''|e^{\im V}e^{-\im \pi l(l+1)/2N} |l\rangle = \delta_{l'l}
\end{equation}
which is equivalent to the time-reflection symmetry of the effective Hamiltonian $H(\beta)=H^*(\beta)$. To construct $\mathcal{T}$ for arbitrary gauges $\beta$, we only need to bring $U$ (or $H$, respectively) back to the $\beta= \sfrac{1}{2}$ gauge. This is accomplished by the gauge transformation $\mathcal{U}(\beta)$ with
\begin{equation}
   \mathcal{U}(\beta)=  e^{-\im \hat{\mathbf{L}}^2\pi(2\beta-1)/(2N)} \with \mathcal{U}^\dagger(\beta)U(\beta)\mathcal{U}(\beta) = U(\sfrac{1}{2}).
\end{equation}
The (anti-unitary) time-reflection operator then takes the form $\mathcal{T}(\beta) = \mathcal{U}(\beta) \, \mathcal{C}\, \mathcal{U}^\dagger(\beta)$ with conjugation $\mathcal{C}$. The time-translation operator then fulfills $\mathcal{T}(\beta)U(\beta)=U^\dagger(\beta)\mathcal{T}(\beta)$, while the Hamiltonian  $[H(\beta), \mathcal{T}(\beta)]=0$. In $k$-space, this leads to 
\begin{equation}
   \mathcal{T}(\beta) H_k(\beta) = H_{-k}^*(\beta)\mathcal{T}(\beta).
\end{equation}
\subsection{$N=3$ case: Dirac cones}
For $N=3$, the Fourier transform of the laser potential becomes 
\begin{equation}
\begin{aligned}
   V(k) &=  \frac{P_1}{2}\begin{pmatrix} 
    1 & e^{-ik}/2 & \sfrac{1}{2}\\
    e^{ik}/2 &1 & e^{-ik}/2 \\
    \sfrac{1}{2} & e^{ik}/2 &1
\end{pmatrix} 
+\frac{P_2}{2}\begin{pmatrix} 
    0 &1 &e^{-ik}\\
    1 &0 &1 \\
    e^{ik} & 1 &0
\end{pmatrix}  
\\ &= \frac{1}{2}\begin{pmatrix} 
    P_1 &P_1e^{-ik}/2+P_2 & P_1/2 + P_2e^{-ik}\\
    P_1 e^{ik}/2+P_2 &P_1 & P_1 e^{-ik}/2+P_2 \\
    P_1/2 +P_2 e^{ik} & P_1 e^{ik}/2+P_2 &P_1
\end{pmatrix}. 
\end{aligned}
\end{equation}
The reflection symmetry $1 \leftrightarrow 3, k \leftrightarrow -k$ is clearly visible here. In Fig.~\ref{fig:fourier1} we show a comparison of the diagonalization in $l-$space and $k-$space. For small to intermediate coupling ($P \sim 5$) the agreement is very good.
If we look at the smallest energy difference $d(P_1,P_2) = \min_{m,n \in \{1,2,3\} }\left( |\epsilon_n - \epsilon_m|_{2\pi}\right )$ with the $2\pi-$periodic distance, \ie 
\begin{equation}
   |x-x'|_{2\pi} =  \min( |x-x'|,|x-x'+2\pi|,|x-x'-2\pi|),
\end{equation}
then we observe that the gap closings occur for $k=0$ and $k=\pi$ on straight lines, see Fig.~\ref{fig:gapclosings}. Note that the linearity of the gap closings can be turned into non linear curves using multiple pulse schemes and more complicated potentials.
\begin{figure}[t]
    \centering
    \includegraphics[width=0.8\textwidth]{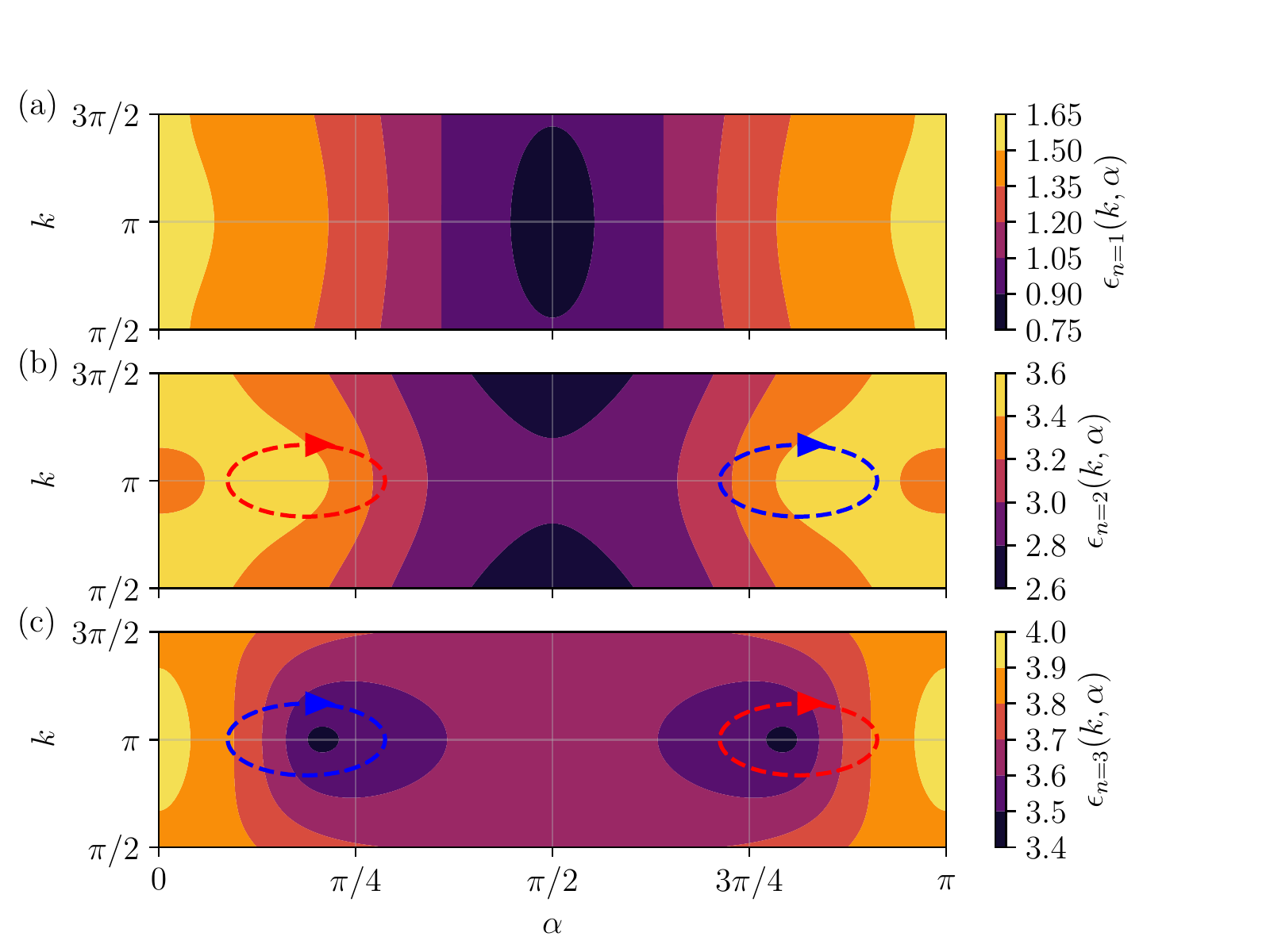}
    \caption{The three energy bands, $\epsilon_n(k,\alpha)$ at $P=2.5$, from (a) lowest to (c) highest. The dashed lines mark exemplary integration contours around the Dirac cones, which provide a way to evaluate the topological charge of the cones, see text. The color of the lines denote the result of the integration with $+\pi$ (red lines) and $-\pi$ (blue lines).}
    \label{fig:vortices}
\end{figure}
Furthermore, we mention in the main text that the topological charge of the Dirac cones can be evaluated numerically. To this end, we combine the two variables as in the main text into a vector $\vec{k}=(k,\alpha),$ and compute the Berry connection, which is well-defined if we choose a gauge such that the eigenstates form a smooth manifold. Choosing a path $\gamma$ around the cones (in the second or third band), the integration of the Berry connection leads to an integer-$\pi$ Berry phase (see Fig.~\ref{fig:vortices} for exemplary contours and the result of the numerical integration~\cite{fukui2005chern})
\begin{equation}
\mathcal{A}_n(\vec{k}) = i\langle \psi_n(\vec{k})|\nabla_\vec{k}\psi_n(\vec{k})\rangle, \quad \int_\gamma \mathcal{A}(\vec{k})_n \d\vec{k}=  \pm \pi.
\end{equation}
\section{Possible experimental realizations of Dirac cones}
\begin{SCfigure}
  \centering
  \caption{The quench protocol of the main text and the corresponding changes of the spectrum in time. For the chosen parameters, $P_1=1.5,P_2=1.6$, the critical value of the parameter is $\alpha_c\approx 0.6221$. The parametrization reads $\alpha(t) =t \cdot (\alpha_c/t_c)$, where we choose $t_c=15$. As for the spectrum in the main text, for $\alpha > \alpha_c$ two edge states emerge in the middle of the two upper bands. They can only be removed by a second cone in between the two bands.}
  \includegraphics[width=0.6\textwidth]%
    {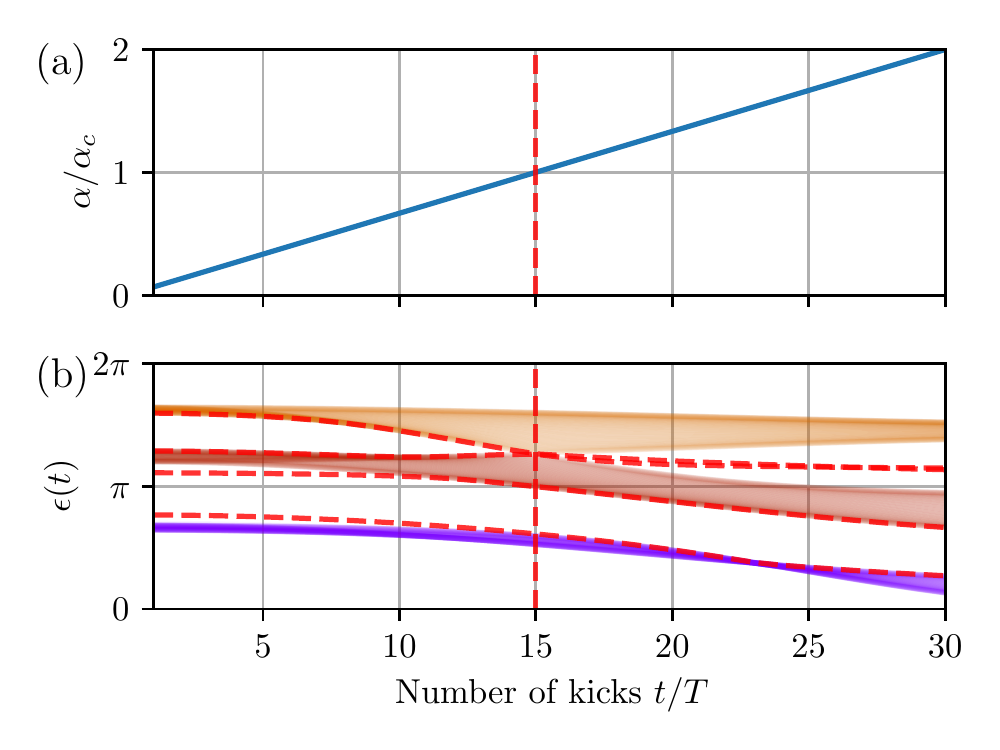}%
    \label{fig:quench1}
\end{SCfigure}

\begin{figure}[t]
    \centering
    \includegraphics[width=0.85\textwidth]{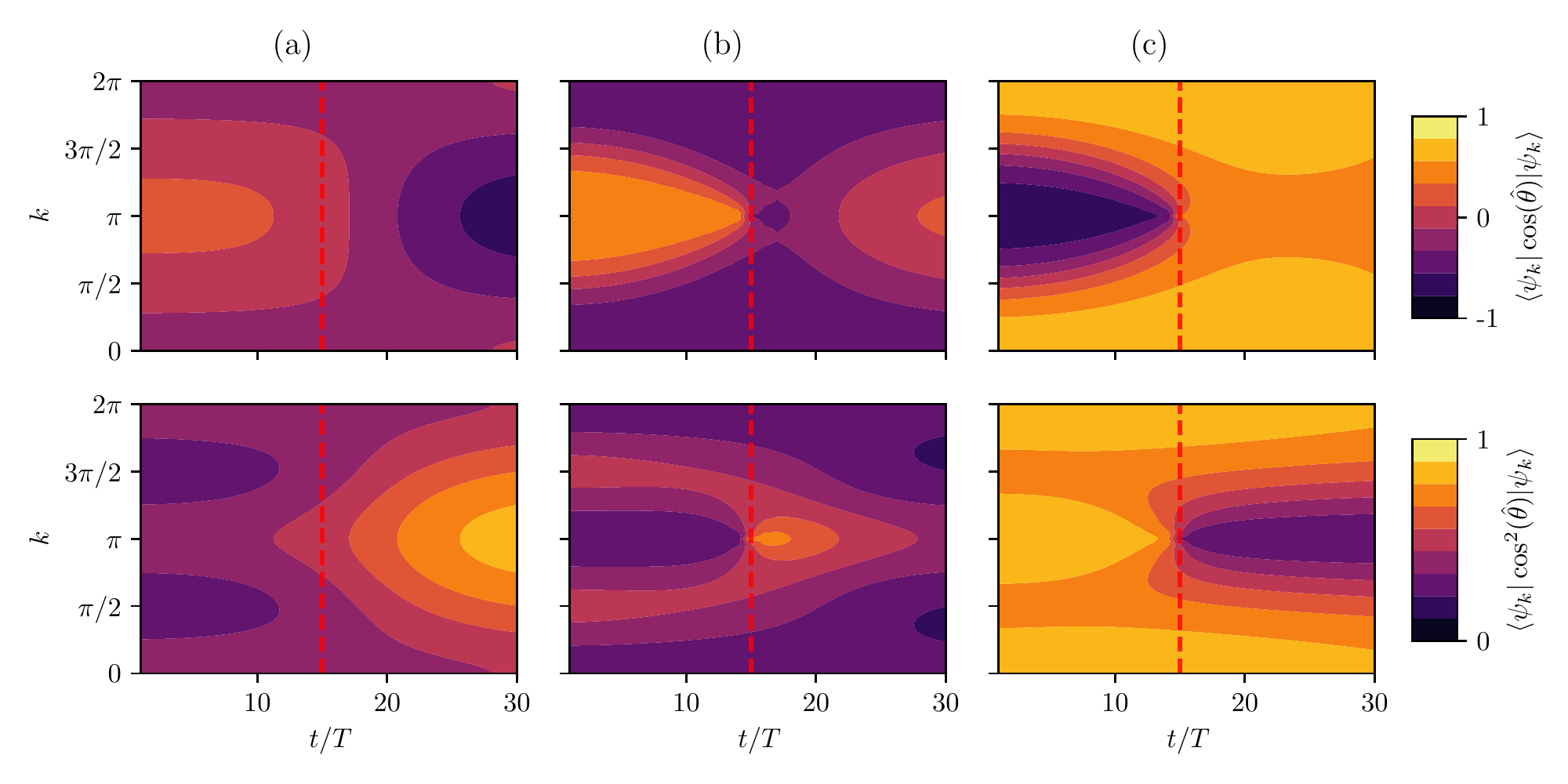}
    \caption{The orientation and alignment signals of the instantaneous eigenstates of the Hamiltonian at a given moment of time. The panels (a),(b),(c) correspond to the first, second and third band, respectively. In the first row we show the expectation value of $\cos(\theta)$ and in the second  of $\cos^2(\theta)$. The red dashed line marks the occurrence of the Dirac cone. The Dirac cone at $t_c=15$ leads to a vortex-like structure in the phases of the eigenstates, which translates into a similar vortex in orientation/ alignment signals. For $k=\pi$, the change is most drastic, e.g.\ $\langle \psi_{k=\pi}|\cos(\hat{\theta})|\phi_{k=\pi}\rangle$ changes sign after passing $t_c$. This figure explains the pronounced changes in the signals of Gaussian wavepackets peaked at $k=\pi$, as shown in the main text, when assuming a quasi-adiabatic time evolution. The change of the behavior of the eigenstates can be explained using the inversion quantum number $n_i=\pm 1$, given by the eigenvalues of the inversion operator defined earlier for inversion-invariant momenta $k=\pi$ and $k=0$. At these momenta, the Hamiltonian becomes block-diagonal. The Dirac cones furnish an exchange of a $n=+1$ state with the $n=-1$ between the bands while maintaining the band structure. This exchange of the states leads to a drastic change of the alignment/orientation signal of the whole band after the cone.}
    \label{fig:quench2}
\end{figure}

\begin{figure}[tb!]
    \centering
    \includegraphics[width=1.0\linewidth]{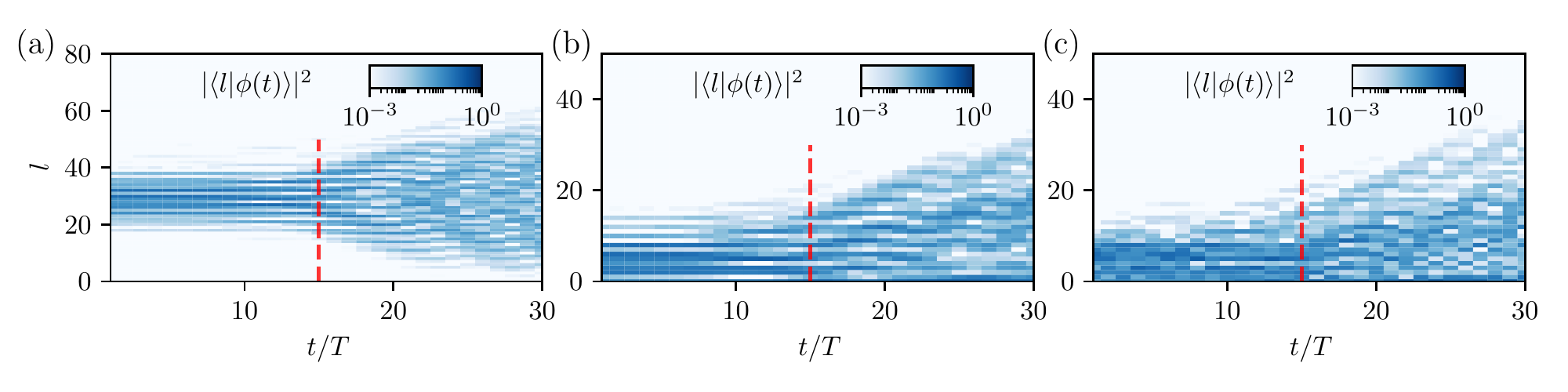}
    \caption{Time evolution of the states discussed in the main text, also see~\eqref{eq:state1}. (a) and (b) show Gaussian states with $\Delta l = \Delta k = 0.5$ with (a) $l_0 = 30$ and (b) $l_0 =0$. (c) shows a ``generic'' state created by a laser pulse starting from $l=0$, $\vert \phi \rangle = e^{-i \Delta t \hat{L}^2}e^{iV(P_1,P_2)}|l=0\rangle$, with $\Delta t\approx 0.911, P_1 \approx 6.67, P_2 \approx 2.67$. These parameters ensure that the state has a large overlap with the third band and is sufficiently occupied around $k=\pi$. We observe that in all three exemplary time evolutions the behavior changes drastically in the vicinity of the Dirac cone. This can be explained by the change of the state due to the band crossing and its spread over different bands, see Fig.~\ref{fig:quench2} and Fig.~\ref{fig:k_quench}. Ultimately, the red line marks a topological transition, where also edge states appear in the middle of the gap from the cone, and they can also be populated.}%
    \label{fig:quench3}
\end{figure}

\begin{figure}[htpb]
    \centering
    \includegraphics[width=0.7\textwidth]{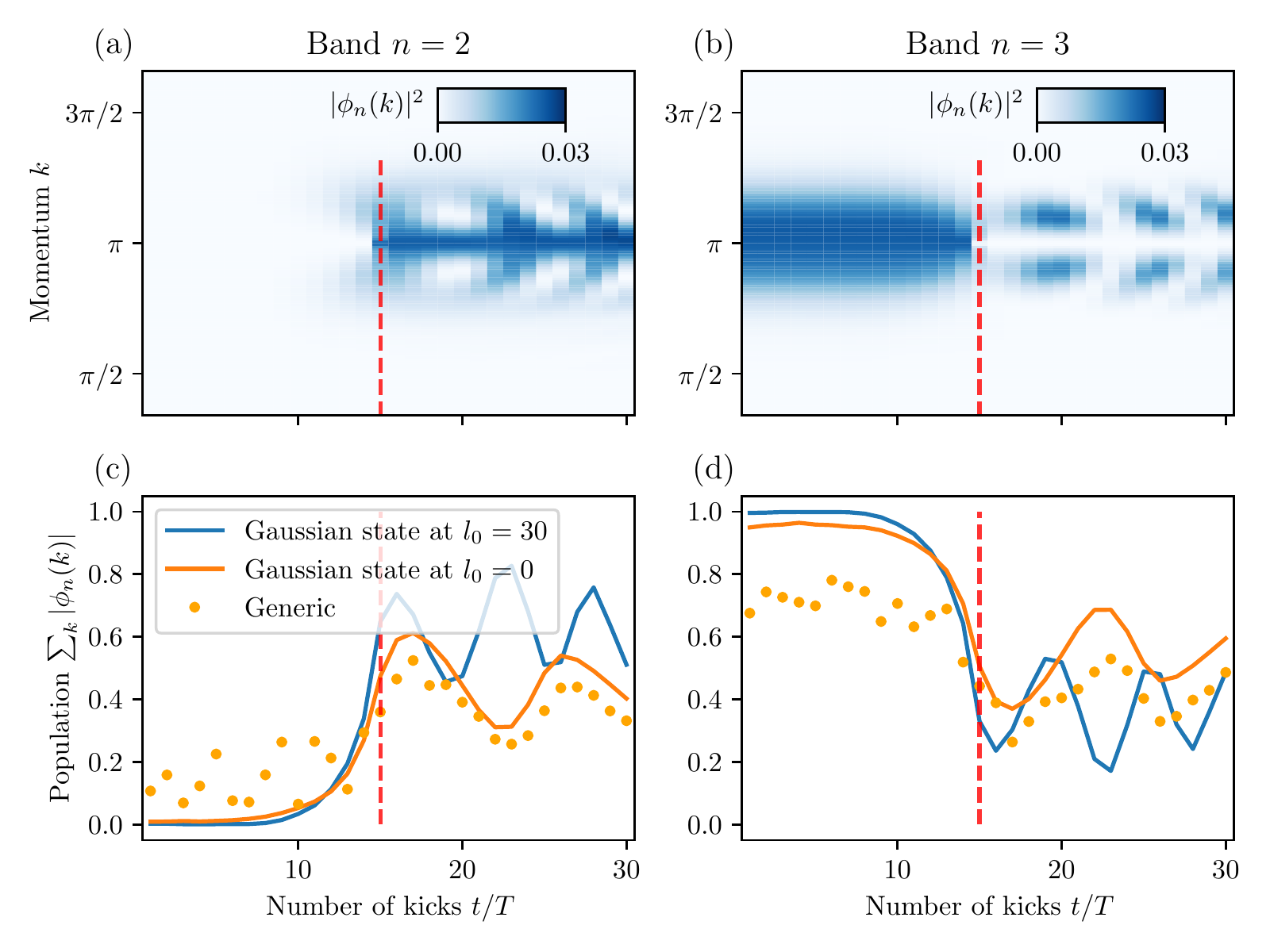}
    \caption{Time-evolution a wavepacket (see Fig.~\ref{fig:quench3}) in $k-$space for the two important bands $n=2$ and $n=3$. The parameters are as before. The Fourier-transform is calculated numerically and cannot capture the complete physics, as it misses finite-size contributions of the lattice. In (a) and (b) we show the time-evolution of the gaussian wavepacket with $l_0=30$. We note that after the Dirac cone (red line) the second band gets occupied. In (c) and (d) we show the total populations in the respective bands. Due to the fast quench over the cone at momenta different from $k=\pi$, there are persistent oscillations in both bands. They originate from the non-adiabatic transitions that occur for small gap differences (in terms of a Landau–Zener transition between two energy states). Only for states that are perfectly occupying $k=\pi$ (and are therefore completely delocalized in angular momentum, such as a angle eigenstate), a clean transition from one band to the other is possible.}
    \label{fig:k_quench}
\end{figure}

\begin{SCfigure}
  \centering
  \caption{The alignment (a) and orientation signal (b), similar to Fig.~2 in the main text, for the gaussian wavepacket at $l_0=30$ and for different critical times $t_c$ (at which the Dirac cone occurs). Here, we set $t_c = t_\mathrm{final} / 2$, \ie the cone is exactly in the middle of the time-evolution. In different colors we show different trajectories for different final times, see colorbar, with the number of kicks ranging from $10 \dots 100$. Clearly, the total number of kicks is not significantly changing the overall behavior. We note that for longer times the transition of band $n=3$ to $n=2$ is smoother and can occur henceforth in a more controlled way.}
\includegraphics[width=0.7\textwidth]{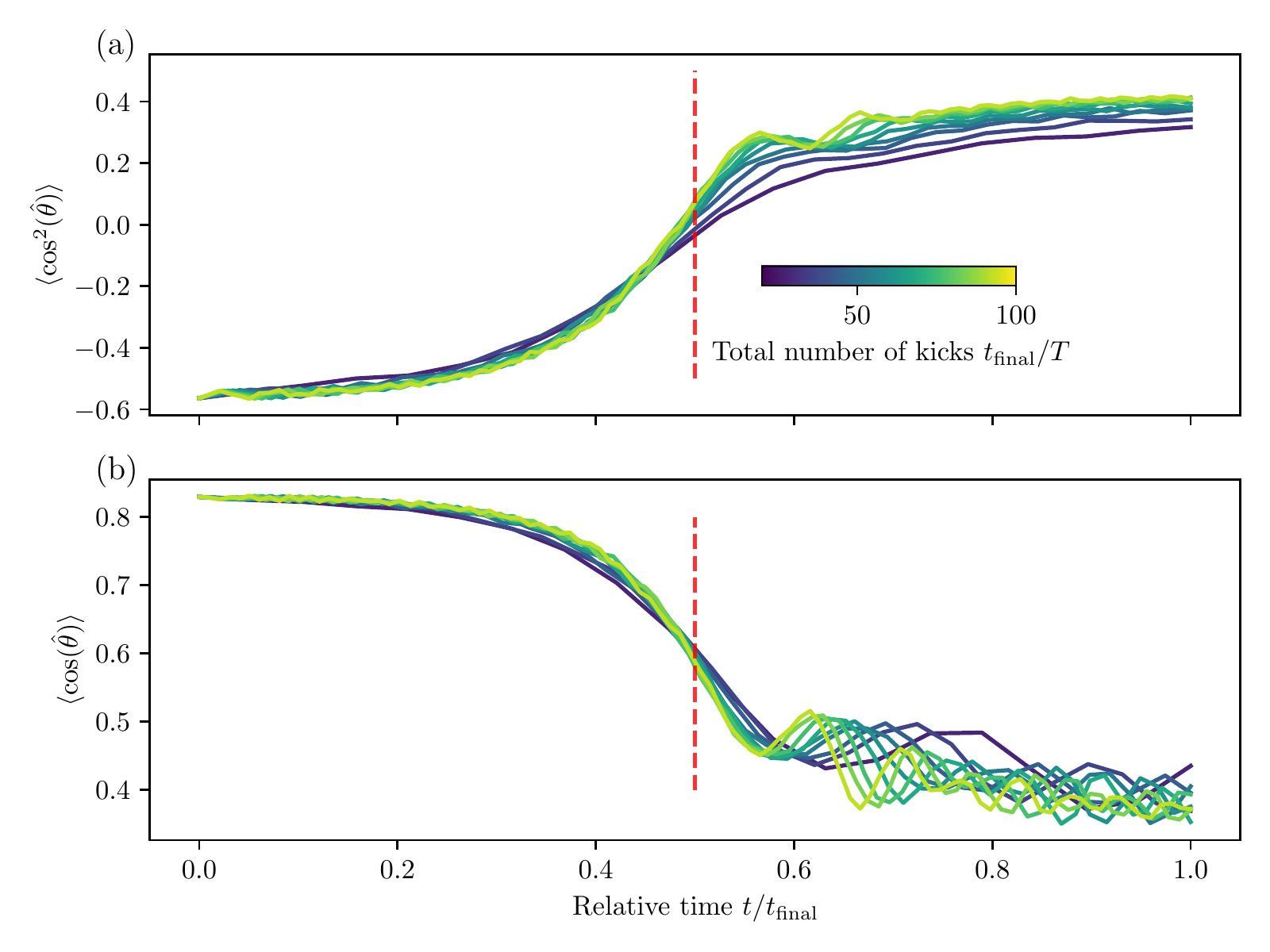}
    \label{fig:differenttc}
\end{SCfigure}

The origin of this behavior can be explained by analyzing the eigenstates of the Hamiltonian. At each moment of time, the eigenstates $\psi(k,\alpha) \in \mathbb{C}^3$ can be obtained by
\begin{equation}
   H(k,\alpha(t)) \psi(k,\alpha(t)) = \epsilon(k,\alpha(t))\psi(k,\alpha(t))
\end{equation}
with $\alpha(t) =t \cdot (\alpha_c/t_c)$ and the critical value $\alpha_c$ with the Dirac cone, see Fig.~\ref{fig:quench1}. In Fig.~\ref{fig:quench2}, we recognize the Dirac cone as a vortex in the expectation value of the eigenstate $\langle \psi_k |\cos^2(\hat{\theta})|\psi_k \rangle$. The Dirac cone causes a drastic change in the orientation and alignment of the eigenstates. A state $\phi(t)$ which consists of a superposition of different momentum eigenstates mimics their evolution accordingly. The drastic change is only observed for states with large occupation at $k=\pi$, since that is the position of the Dirac cone at $t=t_c$. The states for the time-evolution in the main text are obtained as follows. We begin with a Gaussian state in angular momentum space, $\phi_0 \propto e^{-(l-l_0)^2/\Delta l^2}$ and project it into the third band, i.e. $    |\phi_0 \rangle \rightarrow \sum_{i \in \mathcal{N}_3}|\psi_i \rangle \langle \psi_i | \phi_0 \rangle$ with all Floquet states $i \in \mathcal{N}_i$ of the third band. Then, in order to guarantee a peak at $k=\pi$, we convolve this state with a Gaussian distribution centered at $k_0=\pi$, that is  
\begin{equation}
   \tilde{\phi}_k \propto \int_{0}^{2\pi}  e^{-(k-k_0)/2\Delta k^2} (\phi_k \cdot \psi^*_k)\psi_k\,\d k
   \label{eq:state1}
\end{equation}
where we choose $\Delta k = 0.5$. Since this convolution removes overlap with the original Gaussian distribution in $l$-space, it can lead to a delocalization of the former localized state. Clearly, this is a very simple procedure and choosing a more complicated initial state with a well-defined momentum distribution is also possible. Further, in the main text we also show the results for a generic state that was created using only one pulse pulse from $l=0$ (see parameters of this state in the caption of Fig.~\ref{fig:quench3}). As already mentioned in the main text, in order to see a difference in the alignment/orientation signal, one needs to start from a state which is somewhat aligned or oriented, that is one need to choose an appropriate strength in the pulse.\\\\
In Fig.~\ref{fig:quench3} we show the time evolution of the states discussed in the main text. In the beginning of the time evolution, the states do not spread but remain more or less localized, since only a single band is occupied. After crossing the Dirac cone, more bands are occupied (this is partly a non-adiabatic effect) and the wavepacket spreads. However, the change of the alignment/orientation measures is due to the change of the eigenstate, not due to the change of bands, as can be inferred from Fig.~\ref{fig:quench2} (see also caption within, where we explain in detail the role of the quantum number of the inversion symmetry at the transition). The Dirac cone marks the position where these measures suddenly change, and henceforth also the states which depend on them. This implies that by tuning the position of the Dirac cone in terms of the laser intensities, we can also tune when this changes happen for an arbitrary wave function. In Fig.~\ref{fig:k_quench} we show the momentum-dependent time-evolution of a wavepacket, determined by a numerical Fourier transform, and in Fig.~\ref{fig:differenttc} we show the time evolution for different critical times $t_c$ (\ie when the Dirac cone occurs). See the captions within for detailed explanations.

\clearpage

\bibliographystyle{apsrev4-1}
\bibliography{main}